 \documentclass[12pt,preprint]{aastex}

\newcommand{\HESS}{H.E.S.S.}

\newcommand{\naive}{na\"{\i}ve}

\slugcomment{Submitted to the Astrophysical Journal}

\shorttitle{The GeV-TeV Connection in Galactic $\gamma$-ray sources}
  \shortauthors{S.~Funk et al.}

\begin{document}


\title{The GeV-TeV Connection in Galactic $\gamma$-ray sources}


\newcommand{\myemail}{funk@slac.stanford.edu}

\author{S.~Funk\altaffilmark{1},
  O.~Reimer\altaffilmark{2},
  D.~F.~Torres\altaffilmark{3},
  J.~A.~Hinton\altaffilmark{4}
}
\altaffiltext{1}{Kavli Institute for Particle Astrophysics and
  Cosmology (KIPAC), SLAC, CA 94025, USA. funk@slac.stanford.edu}
\altaffiltext{2}{Stanford University, W.~W. Hansen Experimental
  Physics Lab (HEPL) and KIPAC, Stanford, CA 
  94305-4085, USA. olr@stanford.edu}
\altaffiltext{3}{ICREA \& Institut de Ciencies de l'Espai (IEEC-CSIC)
  Campus UAB, Fac. de Ciencies, Torre C5, parell, 2a planta, 08193
  Barcelona,  Spain. dtorres@aliga.ieec.uab.es}
\altaffiltext{4}{School of Physics and Astronomy, University of Leeds,
  Leeds LS2 9JT, UK. jah@ast.leeds.ac.uk} 

\begin{abstract}
Recent observations with atmospheric Cherenkov telescope systems such
as \HESS\ and MAGIC have revealed a large number of new sources of
very-high-energy (VHE) $\gamma$-rays from 100~GeV -- 100~TeV, mostly
concentrated along the Galactic plane. At lower energies (100 MeV --
10 GeV) the satellite-based instrument EGRET revealed a population of
sources clustering along the Galactic Plane. Given their adjacent
energy bands a systematic correlation study between the two source
catalogues seems appropriate. Here, the populations of Galactic
sources in both energy domains are characterised on observational as
well as on phenomenological grounds. Surprisingly few common sources
are found in terms of positional coincidence and spectral
consistency. These common sources and their potential counterparts and
emission mechanisms will be discussed in detail. In cases of detection
only in one energy band, for the first time consistent upper limits in
the other energy band have been derived. The EGRET upper limits are
rather unconstraining due to the sensitivity mismatch to current VHE
instruments. The VHE upper limits put strong constraints on simple
power-law extrapolation of several of the EGRET spectra and thus
strongly suggest cutoffs in the unexplored energy range from 10~GeV --
100~GeV. Physical reasons for the existence of cutoffs and for
differences in the source population at GeV and TeV energies will be
discussed. Finally, predictions will be derived for common GeV--TeV
sources for the upcoming GLAST mission bridging for the first time the
energy gap between current GeV and TeV instruments.
\end{abstract}


\section{Introduction}
In recent years the knowledge of the Galactic VHE $\gamma$-ray sky
above 100~GeV has been greatly improved through the detection and
subsequent study of many sources, mostly by means of ground-based
Imaging Atmospheric Cherenkov telescope systems such as the High
Energy Stereoscopic System (\HESS) or the Major Atmospheric Gamma-ray
Imaging Cherenkov Observatory (MAGIC). Currently known Galactic VHE
$\gamma$-ray emitters include shell-type Supernova remnants
(SNRs)~\citep{HESS1713_II, HESS1713_III, HESSVelaJr}, Pulsar Wind
Nebulae (PWNe)~\citep{HESSMSH, HESS1825II, HESSKooka}, $\gamma$-ray
binaries~\citep{HESSLS5039II, MAGICLSI}, Molecular
clouds~\citep{HESSGCDiffuse} and possibly also clusters of massive
stars~\citep{HESSWesterlund}. These various source classes were
discovered both in pointed observations using \HESS\ and MAGIC as well
as in a systematic survey of the inner Galaxy performed with the
\HESS\ instrument. The highest energy photons detected from these
source classes reach $\sim 100$~TeV~\citep{HESS1713_III}, currently
representing the end of the observable electromagnetic spectrum for
astrophysical objects. It is natural to investigate the relationship
of these TeV sources to sources at lower energies as will be done in
this work focusing on Galactic sources. The closest energy band for
which data exist is that studied by the Energetic Gamma Ray Experiment
Telescope (EGRET) aboard the Compton Gamma-Ray Observatory with an
energetic coverage from 100~MeV -- 10~GeV~\citep{EGRET}.  The GeV sky
has a distinctively different overall appearance compared to TeV
energies. In particular focusing on our Galaxy, the most prominent
feature of the GeV sky is the dominant diffuse emission from cosmic
ray (CR) interactions in the Galaxy, while the TeV sky due to the
steeply falling energy spectrum of the diffuse component is dominated
by individual sources. However, several prominent $\gamma$-ray sources
are known to emit at both GeV and at TeV energies, the Crab Nebula
being the most prominent example~\citep{WhippleCrab,EGRETCrab,
HEGRACrab, HESSCrab, MAGICCrab}.

In this paper the relationship between Galactic EGRET and VHE
$\gamma$-ray sources will be assessed in a systematic way. For cases
with a positional coincidence between a VHE and an EGRET source (in
the following called ``coincident sources'') all currently known
Galactic objects will be considered. For cases in which a source is
detected only in one band -- the ``non-coincident sources'' -- we
focus on the region covered by the \HESS\ Galactic plane survey (GPS)
during 2004 and 2005~\citep{HESSScan, HESSScanII} (Galactic longitude
$\pm 30^{\circ}$, Galactic latitude $\pm 3^{\circ}$) so that a
statistical assessment of the ``non-connection'' can be made. EGRET
was unable to perform detailed studies of the $\gamma$-ray sky above
10~GeV, partly due to back-splash of secondary particles produced by
high-energy $\gamma$-rays causing a self-veto in the monolithic
anti-coincidence detector used to reject charged particles. The
upcoming Gamma Ray Large Area Space Telescope (GLAST) Large Area
Telescope (LAT) will not be strongly affected by this effect since the
anti-coincidence shield was designed in a segmented
fashion~\citep{GLASTACD}. Moreover, the effective area of the
GLAST-LAT will be roughly an order of magnitude larger then that of
EGRET. The GLAST-LAT mission will therefore for the first time fully
bridge the gap between the energy range of EGRET and current VHE
instruments. Part of the study presented here can be seen as
preparatory work for GLAST-LAT studies of sources in the largely
unexplored energy band between 10 and 100~GeV.

\begin{figure}[ht]
\begin{center}
  \includegraphics[width=0.9\textwidth]{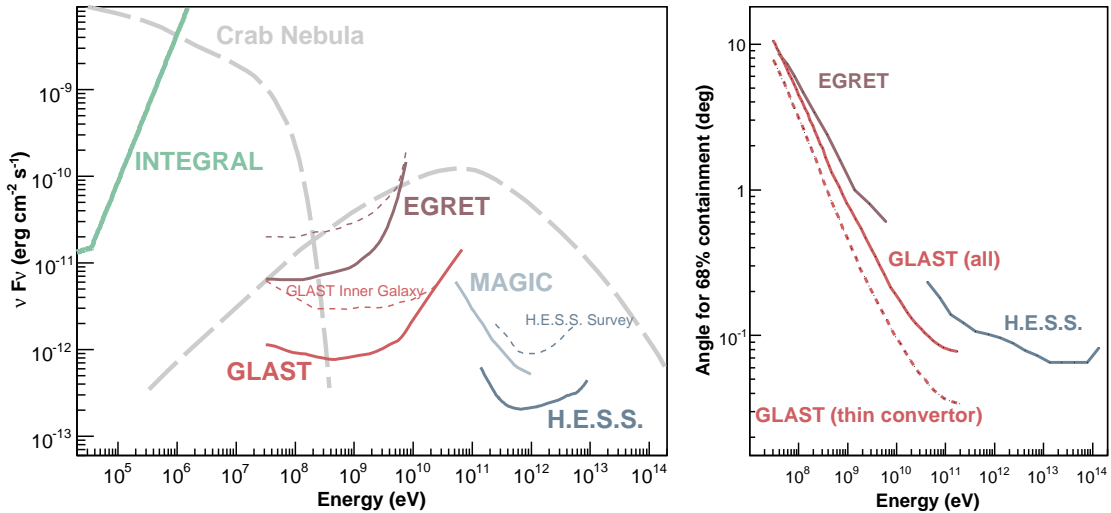} 
\end{center}
\caption{{\bf{Left:}} Integral sensitivities for current, past and
  future $\gamma$-ray instruments (5-$\sigma$ sensitivity for $E>E_0$
  multiplied with $E_0$ assuming a spectrum of $E^{-2}$). The solid
  lines show the nominal instrument sensitivities (for a typical
  observation time as specified below), the dashed curves show the
  actual sensitivities for the Inner Galaxy as appropriate for this
  work. INTEGRAL's (IBIS/ISGRI) sensitivity curve (solid green) shows
  the sensitivity for an observation time of $10^5$s, a typical value
  in the Inner Galaxy. The EGRET curves (brown) are shown for the
  whole lifetime of the mission (periods 1--9) for the Galactic
  anti-centre (solid) which received the largest exposure time and has
  a lower level of diffuse $\gamma$-ray emission than the Inner Galaxy
  and for the position of RX\,J1713.7--3946 (dashed), a typical
  position in the Inner Galaxy dominated by diffuse $\gamma$-ray
  background emission. The GLAST curves in red (taken from
  http://www-glast.slac.stanford.edu/software/IS/glast\_lat\_performance.html)
  show the 1-year sky-survey sensitivity for the Galactic North pole
  -- again a position with low diffuse emission (solid), and for the
  position of RX\,J1713.7--3946 (dashed). The H.E.S.S.\ curves (blue)
  are shown for a 50-hour pointed observation of a point-like source
  (solid) and for a 5-hour observation of a somewhat extended source
  as is typical for the Galactic Plane survey (angular cut of
  $\sqrt{0.05}$). The MAGIC curve (light blue) represents a 50-hour
  observation of a point-source. {\bf{Right:}} Energy-dependence of
  the angular resolution for selected $\gamma$-ray instruments
  expressed by the 68\%-containment radius of the point-spread
  function (PSF). As can be seen, the angular resolution of GLAST
  becomes comparable with current VHE instruments at high energies,
  whilst at the lower energy end GLAST and EGRET have comparable
  resolutions.}\label{fig::Sensitivity}
\end{figure}

\begin{figure}[ht]
\begin{center}
  \includegraphics[width=0.6\textwidth]{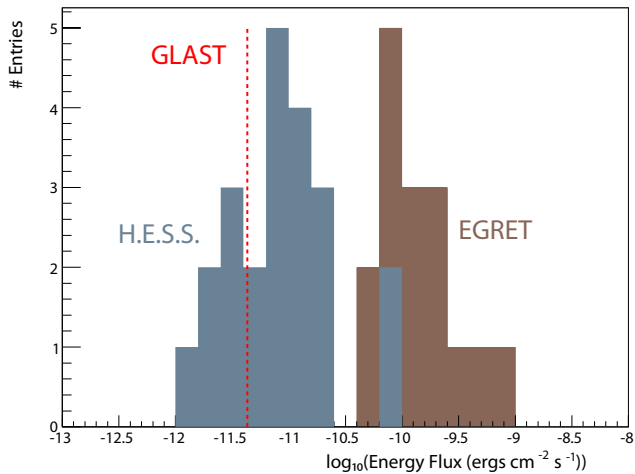} 
\end{center}
\caption{Distribution of integrated energy flux $\nu F \nu$ for
  sources in the Inner Galaxy discussed here. For EGRET the energy
  flux between 1~GeV and 10~GeV, for the \HESS\ sources, the energy
  flux between 1~TeV and 10~TeV is shown. Also shown is the
  sensitivity prediction for the GLAST-LAT for a typical location in
  the Inner Galaxy (l=10, b=0).}\label{fig::EnergyFluxDistribution}
\end{figure}

From the 2004 and 2005 \HESS\ GPS 22 VHE $\gamma$-ray sources have
been reported in the Inner Galaxy. The third EGRET
catalogue~\citep{EGRET} represents the companion to the VHE source
catalogue above an energy threshold of 100 MeV (with peak sensitivity
between 150 and 400 MeV, depending on the $\gamma$-ray source
spectrum). It lists 271 sources, 17 of which are located within the
\HESS\ GPS region. Whilst the EGRET range currently represents the
nearest energy band to VHE $\gamma$-rays, for the very few EGRET
sources detected all the way up to $\sim 10$~GeV, there is still an
unexplored energy band of roughly one decade before the VHE
$\gamma$-ray energy range begins at $\sim 100$~GeV (it should be noted
that EGRET does have some sensitivity beyond 10~GeV:
\citet{EGRET10GeV} reported the detection of $\sim$ 1500 photons above
that energy with 187 of these photons being found within $1^{\circ}$
of a source listed in the third EGRET catalogue). Comparing the
instrumental parameters of VHE instruments and EGRET there is a clear
mismatch both in angular resolution and in sensitivity as can be seen
in Figure~\ref{fig::Sensitivity}. In a $\sim 5$ hour observation (as a
typical value in the GPS region) \HESS\ is a factor of $\sim 50-80$
more sensitive (in terms of energy flux $E^2 dN/dE$) than EGRET above
1~GeV in the Galactic Plane for the exposure accumulated between 1991
and 1995 (corresponding to the third EGRET catalogue). Assuming a
similar energy flux output in the two different bands this mismatch
implies at first sight that \HESS\ sources are not likely to be
visible in the EGRET data set. Conversely (again under the assumption
of equal energy flux output), VHE $\gamma$-ray instruments should be
able to detect the majority of the EGRET sources, as has been
suggested in the past. Figure~\ref{fig::EnergyFluxDistribution}
compares the energy fluxes $\nu F \nu$ for EGRET sources and \HESS\
sources in the inner Galaxy. Clearly, the EGRET sources do not reach
down as low in energy flux as the \HESS\ sources, a picture that will
change once the GLAST-LAT is in orbit as depicted by the GLAST-LAT
sensitivity (dashed line). In reality the \naive\ expectation of equal
energy flux output in the GeV and TeV band can easily be wrong in
Galactic $\gamma$-ray sources for various reasons: EGRET sources may
not emit comparable energy fluxes in the VHE $\gamma$-ray band but
rather exhibit cut-offs or spectral breaks in the energy band between
EGRET and \HESS\ \citep[this is certainly the case for pulsed
magnetospheric emission from pulsars, see for example][]{HESSPulsar}.
Furthermore, \HESS-like instruments are typically only sensitive to
emission on scales smaller than $\sim 1^{\circ}$. If any of the EGRET
sources are extended beyond $1^{\circ}$ without significant
sub-structure on smaller scales (not precluded given the poor
angular resolution of EGRET), current Imaging Cherenkov instruments may not be
able to detect them since these sources would completely fill the
field of view (FoV) and be removed by typical background subtraction
methods ~\citep[see for example][]{BergeBackground}. Given the
upcoming launch of GLAST and the recent \HESS\ survey it seems
timely to study the relationship between GeV and TeV emitting sources
in more detail.

Section~\ref{sec::analysis} describes the data and analysis methods
used in this study, section~\ref{sec::connect} describes the sources
detected in both energy bands, and section~\ref{sec::nonconnect}
focuses on sources detected in only one of the two energy regimes. In
section~\ref{sec::interpretation} astrophysical implications of the
study are discussed.

\section{Analysis methods}
\label{sec::analysis}

For the sources discussed in this study locations and source spectra
in the EGRET band~\citep{EGRET} and in the VHE $\gamma$-ray band are
required. For the inner Galaxy, dedicated upper limits at the specific
position of the $\gamma$-ray sources in the respective other band were
determined.  For the EGRET data these upper limits (at 1~GeV) were
derived at the nominal positions of the \HESS\ sources based on a
reanalysis of the data used for the production of the third EGRET
catalogue, applying the standard EGRET likelihood fitting
technique~\citep{Mattox96}. For the \HESS\ data, 2~$\sigma$ upper
limits at the nominal position of each EGRET source were
estimated. This was done by scaling the flux corresponding to the
\HESS-point-source sensitivity in 25 hours (1\% of the Crab) by the
square-root of the ratio of 25 hours to the published exposure time at
the position of the EGRET source \citep[taken from][]{HESSScanII}.

\subsection{Quantifying Positional Coincidence}

\begin{figure}[ht]
\begin{center}
  \includegraphics[width=0.9\textwidth]{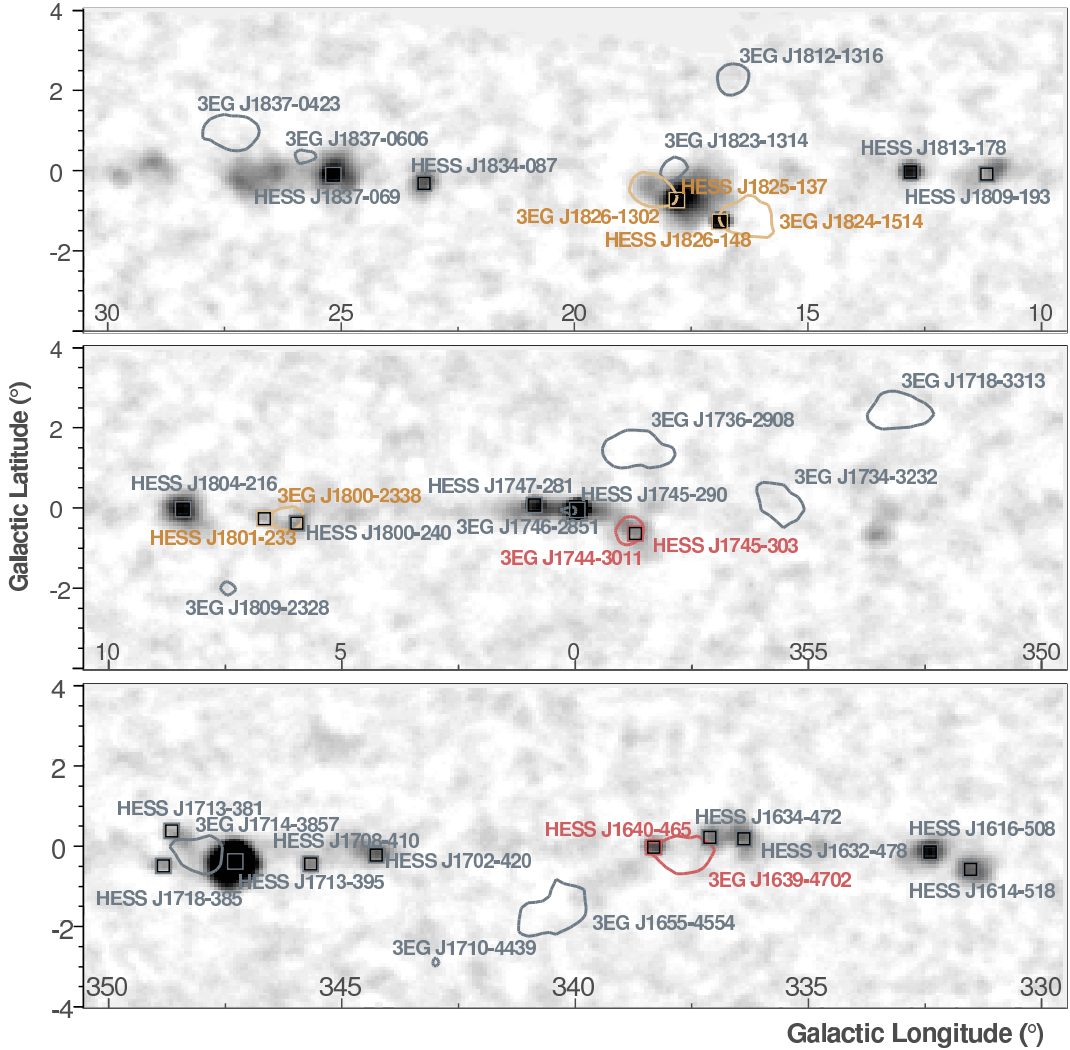} 
\end{center}
\caption{Map of the \HESS\ GPS region taken from \citet{HESSScanII}.
  Published \HESS\ sources are marked as squares. EGRET sources are
  shown with their 95\% positional confidence contours from the
  3EG-catalogue. The red (orange) contours and labels denote those 3EG
  sources for which a \HESS\ source centroid is located within the
  95\% (99\%) confidence contour, the blue contours denote the 95\%
  PUCs for the EGRET sources for which no VHE emission is
  detected.}\label{fig::Map}
\end{figure}

Figure~\ref{fig::Map} shows all \HESS\ and all EGRET sources within
the HESS GPS region. One property of EGRET and VHE $\gamma$-ray
sources becomes immediately apparent: only a minor fraction of the
\HESS\ sources coincide within the considerably larger location
uncertainty contours of EGRET GeV sources. Given the rather poor
angular resolution of EGRET (68\% containment radius of the PSF:
$1.5^{\circ}$ at 1~GeV) coupled with typically rather limited photon
statistic any systematic assessment of positional matches between
EGRET and \HESS\ sources is dominated by the localisation error on the
EGRET source position. The likelihood source position uncertainty
contour (PUC) as given in~\citet{EGRET} have been used to check for
VHE $\gamma$-ray sources within these regions on the sky. While most
of the VHE sources are extended, their extension is rather small on
the scale of the EGRET positional uncertainty and therefore a source
is classified as ``coincident'' if the centre of gravity of the VHE
emission is within the EGRET likelihood PUC. For large sources such as
e.g.\ the SNR RX\,J1713.7--3946 (HESS\,J1713--395)
this approach is clearly an oversimplification, albeit it is the one
used at this stage of this study.

\begin{table}[ht]
  \begin{center}
    \begin{tabular}{c||c|c|c}
      \hline
      EGRET & \multicolumn{3}{c}{VHE $\gamma$-ray} \\
      Source & \multicolumn{3}{c}{Source}  \\
      & Within 68\% & Within 95\% & Within 99\% \\
      & PUC & PUC & PUC \\
      \hline
      & \multicolumn{3}{c}{Within the \HESS\ GPS} \\
      \hline
      3EG\,J1639--4702  & & HESS\,J1640--465 & \\
      3EG\,J1744--3011  & & HESS\,J1745--303 &  \\
      3EG\,J1800--2338  & None & & HESS\,J1800--233 \\
      3EG\,J1824--1514  & & & HESS\,J1826--148 \\
      3EG\,J1826--1302  & & & HESS\,J1825--137 \\
      \hline
      & \multicolumn{3}{c}{Outside the \HESS\ GPS} \\
      \hline
      3EG\,J0241+6103 & & & MAGIC\,J0240+613 \\ 
      3EG\,J0617+2238 & & None & MAGIC\,J0616+225 \\ 
      3EG\,J0634+0521 & & & HESS\,J0632+058 \\  
      3EG\,J1420--6038 & HESS\,J1420--607 &  & \\
      \hline
    \end{tabular}
    \caption{Positionally coincident EGRET and \HESS\ sources within
      our Galaxy for three confidence levels (68\%, 95\%, and 99\%) of
      the positional uncertainty of the EGRET source.}
    \label{tab::position}
  \end{center}
\end{table}

The number of spatially coincident sources depends on the EGRET PUC
chosen in the investigation. For the \HESS\ GPS-region, no VHE
$\gamma$-ray source is located within the 68\% positional confidence
contour of any EGRET source. Relaxing the coincidence criterion, two
VHE $\gamma$-ray sources are found within the 95\%-confidence contour
of EGRET source positions (shown in red in Figure~\ref{fig::Map}) and
an additional three VHE $\gamma$-ray sources are located within the
99\%-confidence contours (shown in orange in
Figure~\ref{fig::Map}). Outside the \HESS\ GPS-region, no systematic
statistical assessment of the non-coincident sources is possible due
to the highly non-uniform exposure of the observations with the
limited-FoV VHE instruments. Nevertheless, is it relevant to note that
four additional coincident sources are found outside the \HESS\
GPS-region within the Galactic plane (defined here as as latitude
range of $\pm 3^{\circ}$): HESS\,J1420--608~\citep{HESSKooka} in the
Kookaburra region is located within the 68\% confidence contour of
3EG\,J1420--6038. The other three coincident Galactic sources are
located within the 99\% confidence contours of EGRET sources. The Crab
Nebula is not listed in Table~\ref{tab::ConnectCases} although it has
been detected by EGRET~\citep{EGRETCrab} as well as by all major VHE
$\gamma$-ray instruments~\citep{WhippleCrab, MilagroCrab, HEGRACrab,
HESSCrab, MAGICCrab}. The reason for this is that in the 3EG catalogue
only the position of the Crab pulsar is given, whereas the PUC of the
off-pulse emission (i.e. the Nebula emission) has not been published
thus far.

These coincident cases are discussed further in
section~\ref{sec::connect}. Table~\ref{tab::position} summarises the
VHE $\gamma$-ray sources located within EGRET confidence contours
inside and outside the \HESS\ GPS region. Within the Galactic Plane
survey region 17.1 square degrees (corresponding to 3\% of the total
GPS region) are covered by the EGRET 95\%-confidence
contours. Randomising the distribution of \HESS\ sources in the region
~\citep[flat in longitude; Gaussian shape in latitude with a mean of
--0.2$^{\circ}$ and a width of 0.34$^{\circ}$ as shown
in][]{HESSScanII} the probability for spatial coincidence between
these two populations can be established. For the 95\% confidence
contours $\sim 1.4$ coincidences between the \HESS-source population
and the EGRET sources is expected. The probability of detecting 2 or
more sources when 1.4 sources are expected by chance is 40\%, i.e. the
positional coincidences could well be expected even if the two
population of sources are not related. Considering the (smaller) 68\%
confidence contours $\sim 0.5$ chance coincidences are expected; the
probability for no coincidence when 0.5 are expected is 60\%. For the
(larger) 99\% confidence contours the picture is similar: $\sim 2.5$
chance coincidences are expected; the probability of finding 5
coincidences when 2.5 are expected is $\sim 9\%$. Summarising these
numbers, it is well possible within the statistics and properties of
the two source classes that all the positional coincidences between
\HESS\ and EGRET sources are chance coincidences. The numbers derived
here do not strongly suggest common sources between the GeV and the
TeV band, although it is (obviously) not precluded that the
coincidences found point to real physical associations.

\subsection{Determining Spectral Compatibility}
\label{sec::spectralmatch}
Besides the test for positional coincidence a test of spectral
compatibility, based on the simple assumption of a spectral
extrapolation by a single power-law between the EGRET and the \HESS\
ranges has been performed. To assess the spectral match the quantity
$\sigma_{\mathrm{comb}}$ has been defined in the following way:
\begin{equation}
  \label{eq::extra}
\sigma_{\mathrm{comb}} = \sqrt{\sigma^2_{\mathrm{3EG}}
+ \sigma^2_{\mathrm{H.E.S.S.}}} 
\end{equation}
To determine $\sigma_{\mathrm{3EG}}$, the spectral index of the EGRET
source has been varied (around the pivot point of the EGRET best fit)
until the extrapolation to 1~TeV matches the \HESS\ flux at that
energy. This pivot point of the EGRET best fit is the energy at which
the error on the index becomes independent from the error on the
normalisation. This resulting index is called
$\Gamma_{\mathrm{match}}$ and
\begin{equation}
  \label{eq::extra2}
\sigma_{\mathrm{3EG}} = (\Gamma_{\mathrm{match}} -
\Gamma_{\mathrm{3EG}})/ (\Delta \Gamma_{\mathrm{3EG}})
\end{equation}
\citep[where $\Gamma_{\mathrm{3EG}}$ and $\Delta
\Gamma_{\mathrm{3EG}}$ is the EGRET index and its error taken
from][]{EGRET}. Consequently, $\sigma_{\mathrm{3EG}}$ is a quantity
that describes by how much the EGRET index has to be altered (with
respect to the error on this index) to match the \HESS\ spectrum at
1~TeV. In the same way $\sigma_{\mathrm{H.E.S.S.}}$, is determined by
changing the \HESS\ spectral index until the flux matches the EGRET
flux at 1~GeV (to avoid biases due to spectral cut-offs at the high
end of the \HESS\ energy range the spectra were fitted only below
1~TeV in cases with clear spectral curvature). The two quantities
$\sigma_{\mathrm{3EG}}$ and $\sigma_{\mathrm{H.E.S.S.}}$ are finally
added in quadrature to yield $\sigma_{\mathrm{comb}}$, describing how
well the two spectra can be extended into each other by a linear
extrapolation.  It should be noted that for the procedure described
here, only the statistical (not the systematic) errors on the spectral
indices are taken into account. For cases with a source detection only
in one band, the same procedure can be applied using the upper limit
in the other band (with the obvious difference that only the
extrapolation from the source spectrum onto the upper limit can be
performed, not the other way around). For cases in which the power-law
extrapolation with the nominal source photon index turns out to be
lower -- and therefore non-constraining -- to the upper limit the
corresponding measure $\sigma_{\mathrm{3EG}}$ or
$\sigma_{\mathrm{H.E.S.S.}}$ is set to zero (i.e. the spectra are
compatible). In several (but not the majority of) cases the EGRET
spectrum can be preferentially fit by a higher order spectral shape
(e.g.\ an exponential cutoff or a broken power-law) as will be
discussed in section~\ref{sec::nonconnect}.

\section{VHE $\gamma$-ray sources with EGRET counterparts}
\label{sec::connect}

Only a few coincident sources between the GeV and the TeV band have
been reported so far. The VHE $\gamma$-ray sources that positionally
coincide with EGRET sources are summarised in
Table~\ref{tab::position}. Whilst the positional coincidences between
EGRET and VHE $\gamma$-ray sources might all be chance coincidences as
shown in the previous section, in the following all positional
coincidences within the 99\% EGRET PUCs will be considered. Some of
the properties of the sources and their respective source classes will
be discussed along with an investigation on their spectral
compatibility as introduced in the previous section.

\subsection{Source Classes}

For EGRET sources in the Galactic plane, the only firm identifications
with sources at other wavelengths are for pulsars, based on matching
radio or X-ray periodicity~\citep{Thomp94}. For many of the remaining
Galactic EGRET sources, counterparts have been suggested, but the
angular resolution of the instrument and the strong diffuse
$\gamma$-ray background in the Galactic plane prevented unambiguous
identifications. In VHE $\gamma$-rays, several source classes have
been firmly identified as has been discussed e.g.\ in
~\citet{FunkBarcelona}, based on matching morphology, positional
coincidence or periodicity. However, the majority of Galactic VHE
$\gamma$-ray sources also remain unidentified.
Table~\ref{tab::ConnectCases} summarises potential counterparts of VHE
sources in the coincident cases. While some of these identifications
are rather solid (as e.g.\ in the case of the $\gamma$-ray binaries
LS\,5039~\citep{HESSLS5039II} and LSI\,+61\,303~\citep{MAGICLSI}), in
most of the other cases the identification of (even) the VHE
$\gamma$-ray sources (with relatively small PUC of $\sim 1\arcmin$)
lack any evidence of association beyond positional coincidence.  In
the cases where a firm identification exists, the VHE $\gamma$-ray
source can be used to shed light on the nature of GeV source, assuming
a physical relationship as shown for the Kookaburra
region~\citep{Reimer}. Such studies demonstrate that observations with
VHE $\gamma$-ray instruments can provide templates necessary to pin
down the nature of unidentified EGRET $\gamma$-ray sources with
suggestive but unproven counterparts. With the upcoming advent of the
GLAST-LAT instrument this approach will become very useful for
associating the GeV emission as measured by a large-aperture
space-based $\gamma$-ray instrument with narrow FoV but superior
spatial resolution observations of ground-based VHE $\gamma$-ray
instruments. Provided that the physical associations discussed in this
section and shown in Table~\ref{tab::ConnectCases} are confirmed (as
e.g., through more sensitive measurements with GLAST-LAT), three
long-suspected classes of Galactic GeV sources (SNRs, PWNe and Binary
systems) could finally be conclusively established. In the following
these different source classes will be briefly discussed in the
context of this study.

\begin{table}[ht]
  \begin{center}
    \begin{tabular}{l c | l}
      \hline EGRET source & VHE $\gamma$-ray source & Potential
      Counterpart\\ \hline \multicolumn{3}{c}{Within the \HESS\ GPS}
      \\ \hline 3EG\,J1639--4702 & HESS\,J1640--465 & G338.3--0.0
      (SNR/PWN)\\ 3EG\,J1744--3011 & HESS\,J1745--303 & \\
      3EG\,J1800--2338 & HESS\,J1801--233 & W28 (SNR)\\
      3EG\,J1826--1302 & HESS\,J1825--137 & G18.0--0.7 (PWN)\\
      3EG\,J1824--1514 & HESS\,J1826--148 & LS\,5039 (Binary)\\ \hline
      \multicolumn{3}{c}{Outside the \HESS\ GPS} \\ \hline
      3EG\,J0241+6103 & MAGIC\,J0240+613 & LSI\,+61\,303 (Binary)\\
      3EG\,J0617+2238 & MAGIC\,J0616+225 & IC443 (SNR/PWN)\\
      3EG\,J0634+0521 & HESS\,J0632+058 & Monoceros \\
      3EG\,J1420--6038 & HESS\,J1420--607 & Kookaburra (PWN)\\ \hline
    \end{tabular}
    \caption{Coincident sources and potential counterparts to the VHE
    $\gamma$-ray sources (and hence also to the associated EGRET
    sources). The counterparts are classified into the source classes
    shell-type SNRs, PWNe and $\gamma$-ray binaries.}
  \label{tab::ConnectCases}
  \end{center}
\end{table}

\subsubsection{Pulsar wind nebulae} 
PWNe are currently the most abundant class amongst the identified
Galactic VHE $\gamma$-ray sources, it is not therefore surprising that
PWN are found as potential counterparts to the coincident sources. The
first example for a coincident PWN is HESS\,J1825--137 -- located
within the 99\% confidence region of 3EG\,1826--1302. This
source~\citep{HESS1825II} is currently the best-known example for an
offset $\gamma$-ray PWN and as such represents a prototype for a new
class of $\gamma$-ray sources. HESS\,J1825--137 shows a steepening of
the energy spectrum with increasing distance from the central
pulsar. This property, as well as the observed difference in size
between the VHE $\gamma$-ray emitting region and the X-ray PWN
associated with the pulsar PSR\,B1823--13~\citep{Gaensler} can be
naturally explained by different cooling timescales for the radiating
electrons of different energies. In this regard it will be important
to study this region with the GLAST-LAT in the GeV band to confirm (or
refute) this picture. Another example of VHE $\gamma$-ray PWN is
HESS\,J1420--607, one of two likely PWN in the previously discussed
Kookaburra region, which is located within the 68\% confidence region
of 3EG\,J1420--6038. The Crab Nebula is not listed in
Table~\ref{tab::ConnectCases} although being a prominent GeV and TeV
source because no position for the Crab off-pulse (nebular) emission
has been published for the GeV data.  For previously unidentified
sources such as HESS\,J1640--465 (G338.3--0.0) an association with the
X-ray PWN~\citep{Funk1640} is suggestive but not firmly established at
this point.

\subsubsection{Shell-type Supernova remnants} 
Shell-type SNRs constitute another prominent class of VHE $\gamma$-ray
sources. However, the two most prominent VHE $\gamma$-ray shell-type
SNRs RX\,J1713.7--3946 and RX\,J0852.0--4622 (Vela Jr.) are not
prominent GeV emitters even though they are (up to now) the brightest
steady VHE $\gamma$-ray sources in the sky after the Crab Nebula. Also
Cas~A and RCW~86 have been reported as VHE $\gamma$-ray
sources~\citep{CasAHegra, MagicCasA, RCW86HESS} but have not been
detected by EGRET.  \citet{Sturner, Esposito, Romero, DiegoSNR}
assessed the relationship between unidentified EGRET sources at low
Galactic latitude and SNRs and found a statistically significant
correlation between the two populations at the 4--5 $\sigma$ level,
were however not able to firmly and uniquely identify individual SNRs
as EGRET sources. The GLAST-LAT will shed more light on the GeV
emission in this source as well as in the whole population of Galactic
SNRs. By measuring shape and level of the high-energy
$\gamma$-ray emission the GLAST-LAT might allow for a distinction
between hadronic and leptonic emission models as discussed in
section~\ref{sec::interpretation}. Other potential shell-type SNR
counterparts related to this analysis are W\,28 (HESS\,J1801--233 and
3EG\,J1800--2338), IC443/MAGIC\,J0616+225~\citep{MAGICIC443}, and the
Monoceros Loop SNR (HESS\,J0632+058 and
3EG\,J0634+0521)~\citep{HESSMonoceros}, although in particular in the
latter case, the morphology of the VHE $\gamma$-ray source does not
lend support to an association with the SNR shell.

\subsubsection{$\gamma$-ray binaries} 
Three binary systems: PSR\,B1259--63/SS\,2883, LS\,5039 and
LSI\,+61\,303, have now been established as VHE $\gamma$-ray
sources~\citep{HESS1259, HESSLS5039II, MAGICLSI, VeritasLSI}.  The
latter two of these objects have long been considered as likely
counterparts to EGRET sources~\citep{Kniffen_LSI, Tavani_LSI, EGRET,
Paredes_LS5039}, however, a definitive identification could not be
achieved in the GeV waveband so far. The VHE $\gamma$-ray emission is
undoubtedly related to the binary system (as e.g.\ in LS\,5039
established through the detection of characteristic periodicity,
matching the orbital period of the binary system), strengthening the
case that the GeV emission is also associated to these binaries.
Recently, the MAGIC collaboration presented evidence for VHE
$\gamma$-ray emission from the black-hole X-ray binary Cyg\,X-1,
during a flaring state in X-rays~\citep{MAGICCygX}.  There is no
evidence so far for GeV emission from this object.

\subsection{Spectral compatibility}

\begin{figure}[ht]
\begin{center}
\noindent
\includegraphics[width=0.95\textwidth]{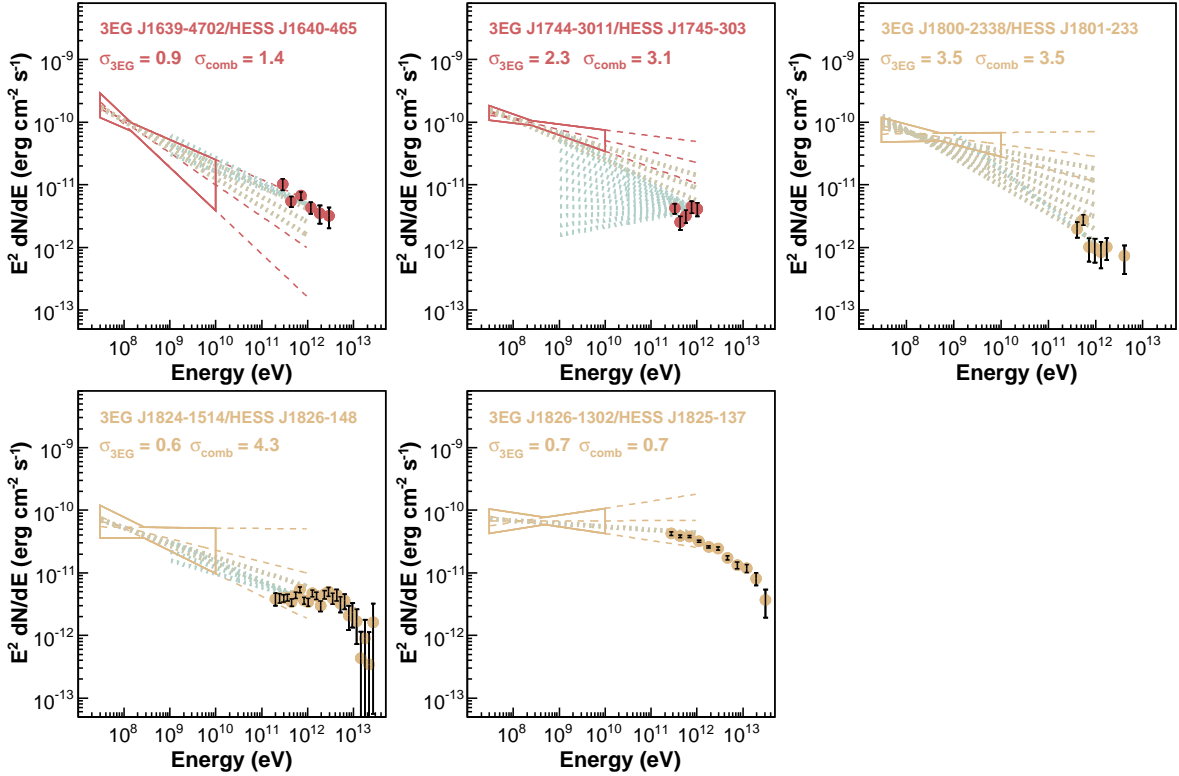}
\end{center}
\caption{Spectra for the positionally coincident EGRET and \HESS\
  sources within the \HESS\ GPS region. Sources for which the
  H.E.S.S.\ source is located within the 95\% confidence level are
  shown in red, whereas those within the 99\% confidence contour (as
  give in Table~\ref{tab::position}) are shown in orange. The EGRET
  ``butterfly'' is determined from the 3EG catalogue~\citep{EGRET},
  the \HESS\ spectral points are taken from the respective
  publication.  For HESS\,J1826--148 and HESS\,J1825--137, which have
  significantly curved TeV spectra, only the spectral points below
  1~TeV have been fitted and used for the extrapolation. Large values
  of $\sigma_{\mathrm{comb}}$ indicate mismatches between the spectra
  at GeV and at TeV energies.}
\label{fig::SpectraConnect}
\end{figure}

As described in section~\ref{sec::spectralmatch}, a test for
compatibility between EGRET and \HESS\ energy spectra based on a
single power-law extrapolation has been performed, calculating for
each of the coincident cases in the \HESS\ GPS region a measure of
spectral mismatch: $\sigma_{\mathrm{comb}}$.
Figure~\ref{fig::SpectraConnect} shows the result of these
extrapolations. The values for the spectral compatibility parameter
$\sigma_{\mathrm{comb}}$ are rather small.  The largest value,
potentially indicative of a spectral mismatch, is found for the case
of the $\gamma$-ray binary association LS\,5039 (3EG\,J1824--1514 and
HESS\,J1826--148). However, this value is completely dominated by the
small statistical error on the \HESS\ power-law fit below 1~TeV (error
on the photon index: $\Delta \Gamma_{\mathrm{stat}} = \pm
0.04$). Taking a typical \HESS\ systematic error of $\Delta
\Gamma_{\mathrm{sys}} = \pm 0.2$ on the determination of the photon
index $\Gamma$ into account, the GeV and TeV energy spectra in this
source match well. Figure~\ref{fig::SpectraConnect} therefore suggests
that the energy spectra of sources that show a spatial association can
generally be rather well described by a single power-law description
across the entire energy range from 0.1~GeV to 1~TeV.

\begin{figure}[ht]
  \begin{center}
    \noindent
    \includegraphics[width=0.7\textwidth]{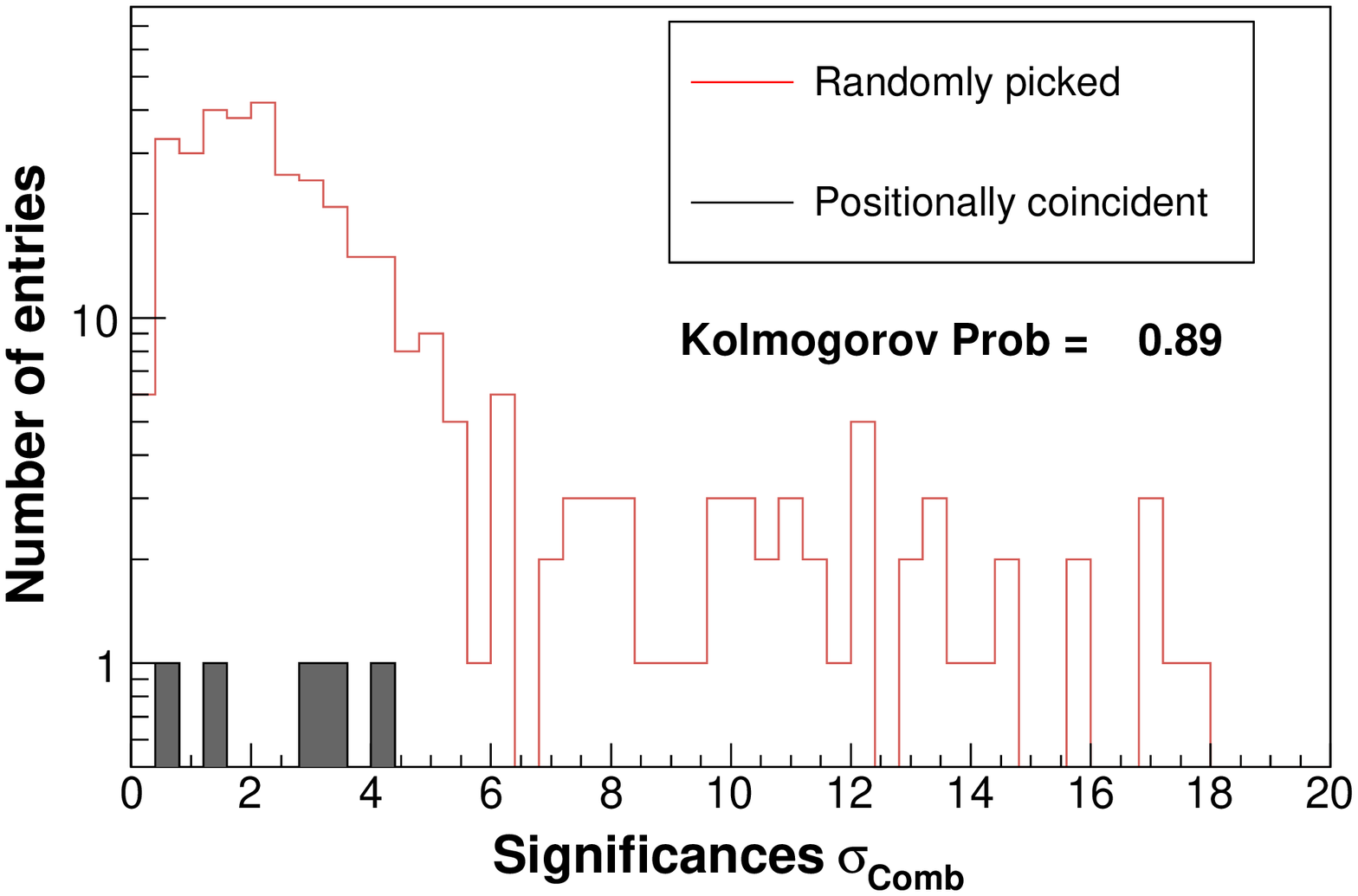} 
  \end{center}
  \caption{Distribution of $\sigma_{\mathrm{comb}}$. The red histogram
    shows the distribution for the spectral consistency parameter
    $\sigma_{\mathrm{comb}}$ of all possible combinations of EGRET
    sources with \HESS\ sources within the GPS region. The
    black histogram shows the same distribution for the 5 cases of
    positional coincidences.}\label{fig::ChanceCoincidence}
\end{figure}

To estimate the chance coincidence of a spectral consistency the
spectra of all 17 EGRET sources and of all 22 \HESS\ sources in the
\HESS\ GPS region have been interchanged and ``connected'' to each
other (i.e.\ the spectral compatibility of each \HESS\ source has been
determined for each EGRET source -- regardless of positional
coincidence).  The resulting distribution of $\sigma_{\mathrm{comb}}$
can be interpreted as the probability density function for
$\sigma_{\mathrm{comb}}$ for randomly selected \HESS\ and EGRET
sources and is shown in Figure~\ref{fig::ChanceCoincidence} as a red
histogram. This distribution should be compared to the measured
distribution of $\sigma_{\mathrm{comb}}$ for positionally coincident
pairs (black histogram). Even though the distribution for the
scrambled sources shows a tail to large values of
$\sigma_{\mathrm{comb}}$, most random pairings result in values of
$\sigma_{\mathrm{comb}}<5$.  A Kolmogorov test yields a probability of
89\% that the two distributions are based on a common underlying
distribution. Thus a spectral compatibility based on a power-law
extrapolation of a typical (randomly picked) EGRET and a typical
\HESS\ source is expected to occur by chance even in the absence of a
physical association.  This is perhaps not surprising, given that both
EGRET as and \HESS\ spectra have typical photon indices of $\sim 2.3$
and that \HESS\ measurements occur $\sim 4$ orders of magnitude higher
in energy with \HESS\ being 1--2 orders of magnitude more sensitive.

\section{Inner Galaxy $\gamma$-ray sources detected in only one band}
\label{sec::nonconnect}

In this section the remainder (and majority) of sources in the \HESS\
GPS region will be discussed. These are the sources which do not have
a counterpart in the neighbouring energy band. In
Section~\ref{sec::nonconnect::EGRET} EGRET sources without a VHE
$\gamma$-ray counterpart will be discussed,
section~\ref{sec::nonconnect::VHE} investigates VHE $\gamma$-ray
sources without an EGRET counterpart.

\subsection{EGRET sources without a VHE $\gamma$-ray counterpart}
\label{sec::nonconnect::EGRET}

\begin{table}[t]
  \begin{center}
    \begin{tabular}{c|c |r}
      \hline
      EGRET & H.E.S.S. & $\sigma_{\mathrm{comb}}$\\
      Source & Upper Limit &  \\
      & ($10^{-12}$ ergs cm$^{-2}$ s$^{-1}$) & \\
      \hline
      3EG\,J1655-4554 & 0.4 & 1.3\\
      3EG\,J1710-4439 & 1.5 & 16.3\\
      3EG\,J1714-3857 & 0.2 & 1.5\\
      3EG\,J1718-3313 & 1.0 & 0\\
      3EG\,J1734-3232 & 0.6 & 1.4\\
      3EG\,J1736-2908 & 0.3 & 3.5\\
      3EG\,J1746-2851 & 0.2 & 15.7\\
      3EG\,J1809-2328 & 0.5 & 6.4\\
      3EG\,J1812-1316 & 2.1 & 1.0\\
      3EG\,J1823-1314 & 0.4 & 0\\
      3EG\,J1837-0423 & 0.6 & 0\\
      3EG\,J1837-0606 & 0.4 & 5.5\\
      \hline
    \end{tabular}
    \caption{EGRET sources without a VHE $\gamma$-ray counterpart in
      the \HESS\ GPS region. The \HESS\ differential upper limits
      (2~$\sigma$) at 1~TeV for a point-source analysis, are derived
      from the \HESS\ 2004--2005 exposure at the nominal EGRET
      position as described in the text (under the assumption of a
      photon index of 2.6).}
      \label{tab::nonconnect}
  \end{center}
\end{table}

\begin{figure}[ht]
  \begin{center}
    \noindent
    \includegraphics[width=0.7\textwidth]{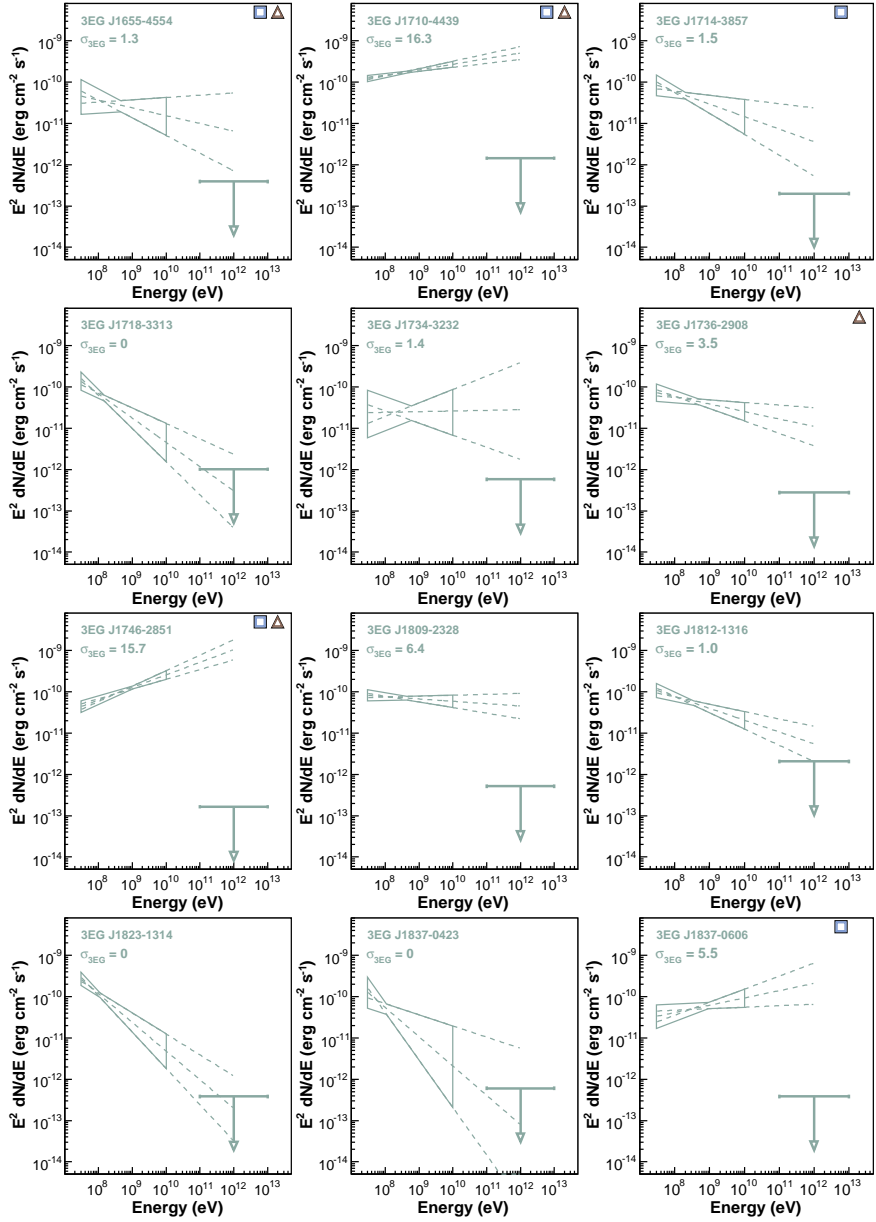} 
  \end{center}
  \caption{SED of the EGRET source for which no VHE $\gamma$-ray
    source was found within the 99\% confidence contour. Sources
    marked with a square show $\gamma$-ray emission above 10~GeV in
    the EGRET data as reported by~\citet{EGRET10GeV}, for sources
    marked with a triangle the EGRET data are better described by
    either a broken power-law or a power-law with an exponential
    cutoff as shown in Figure~\ref{fig::OtherFits}.}
  \label{fig::EGRETnonconnect}
\end{figure}

Here those EGRET sources are addressed with a 99\%-confidence centroid
position region which does not contain a reported VHE $\gamma$-ray
source centroid.  This sample consist of 12 EGRET detections, with E
$> 100$ MeV fluxes ranging between 0.4 and 3.1 $\times
10^{-6}$\,cm$^{-2}$\,s$^{-1}$ and photon indices of the power-law fits
between $\sim$1.75 and 3.2. For these 12 EGRET sources, 2~$\sigma$
upper limits on the VHE flux at 1~TeV for the nominal EGRET position
were determined. This was done by scaling the \HESS\ sensitivity for a
5~$\sigma$ point source detection (1\% of the Crab in 25 h under the
assumption of a photon index of 2.6) to the actual exposures as
published for the \HESS\ GPS region~\citep{HESSScanII}.  As described
previously, the spectral compatibility parameter
$\sigma_{\mathrm{3EG}}$ was determined according to
Equation~\ref{eq::extra2}. For cases in which the EGRET extrapolation
with the nominal 3EG photon index undershoots the \HESS\ upper limit,
$\sigma_{\mathrm{3EG}}$ is set to zero. The resulting plots are shown
in Figure~\ref{fig::EGRETnonconnect}. For Gaussian errors
$\sigma_{\mathrm{3EG}}$ represents the probability that the true GeV
spectrum would pass through the HESS upper limit.

In seven of the twelve cases the \HESS\ upper limit does not impose a
strong constraint on an extrapolation of the EGRET spectrum
($\sigma_{\mathrm{3EG}} < 1.5$).  For the remaining five sources a
\HESS\ detection would have been expected, based on a \naive\
power-law extrapolation. In particular extrapolations for four of the
EGRET sources exhibiting a hard energy spectrum (3EG\,J1710--4439,
3EG\,J1746--2851, 3EG\,J1809--2328, and 3EG\,J1837--0606) are
incompatible with the \HESS\ upper limits at levels exceeding
$\sigma_{\mathrm{3EG}}>5$. For these cases the VHE $\gamma$-ray data
strongly suggest a spectral turnover (cutoff or break) well below the
\HESS\ range. Such behaviour is not surprising for some Galactic
source classes. For the EGRET-detected pulsars a cutoff in the energy
spectrum is seen in many sources \emph{within} the EGRET energy regime
(and therefore clearly well below the VHE range). Indeed, for three
out of the four EGRET sources for which a spectral change is implied
by the \HESS\ non-detection, a pulsar association has been proposed:
3EG\,J1710--4439 was unambiguously identified with
PSR\,1706--44~\citep{Thomp94}, 3EG\,J1809--2328 was proposed to be of
PWN nature~\citep{Braje02}, and 3EG\,J1837--0606 was suggested as the
counterpart of PSR\,J1837--0604~\citep{DAmico01}. The remaining source
in the sample for which the spectral extrapolation of the EGRET source
is constrained by the \HESS\ upper limit, is the Galactic Centre
source 3EG\,J1746--2851. This object is extremely interesting and
important, and may be related to the TeV emission detected in this
region, however, a proper discussion falls beyond the scope of
this paper.

It is interesting to note that an analysis of the EGRET data above
10~GeV~\citep{EGRET10GeV} found eleven EGRET sources with evidence for
emission above 10~GeV (at a level of less than 10\% probability that
the number of photons seen is a fluctuation of the diffuse background
emission). Five of these sources are located in the \HESS\ GPS
region. These sources are 3EG\,J1655--4554, 3EG\,1710--4439
(PSR\,B1706--44, with a 6.1$\sigma$ detection significance above
10~GeV) 3EG\,J1714--3857, 3EG\,J1746--2851, and 3EG\,J1837--0606 (all
marked with a white-and-blue square in
Figure~\ref{fig::EGRETnonconnect}). Interestingly, all of these
sources belong to the class of non-coincident sources, i.e. have no
counterpart at VHE $\gamma$-ray energies. The characteristic cut-off
energies of these sources are therefore likely confined to the region
below $\sim 100$ GeV. This emphatically emphasises the existence of
cutoffs within the energetic gap left between the end of the EGRET
measurements and the onset of the \HESS\ and MAGIC observations.

To further investigate the cutoff hypothesis a spectral analysis of
the EGRET energy spectra has been performed by means of higher order
representations, as has been reported by ~\citet{Bertsch00,
ReimerBertsch01}. The EGRET spectra were fitted with a broken
power-law and with a power-law with an exponential cutoff:

\begin{equation}
\frac{\partial{J}}{\partial{E}}(E,K,\lambda _{1}, \lambda _{2}) = \left\{
\begin{array}{ll} K \left(\frac{E}{1 \mathrm{GeV}}\right)^{-\lambda _{1}} (E
\leq 1 \mathrm{GeV})  \nonumber\\
K \left(\frac{E}{1 \mathrm{GeV}}\right)^{-\lambda _{2}} (E \geq 1
\mathrm{GeV})
\end{array} \right.
\end{equation}

\begin{equation}
\frac{\partial{J}}{\partial{E}}(E,K,\lambda,E_{c}) = K \left(\frac{E}{300
\mathrm{MeV}}\right)^{-\lambda} exp\left(-\frac{E}{E_{c}}\right)
\end{equation}

\begin{figure}[ht]
  \begin{center}
    \noindent
    \includegraphics[width=0.57\textwidth]{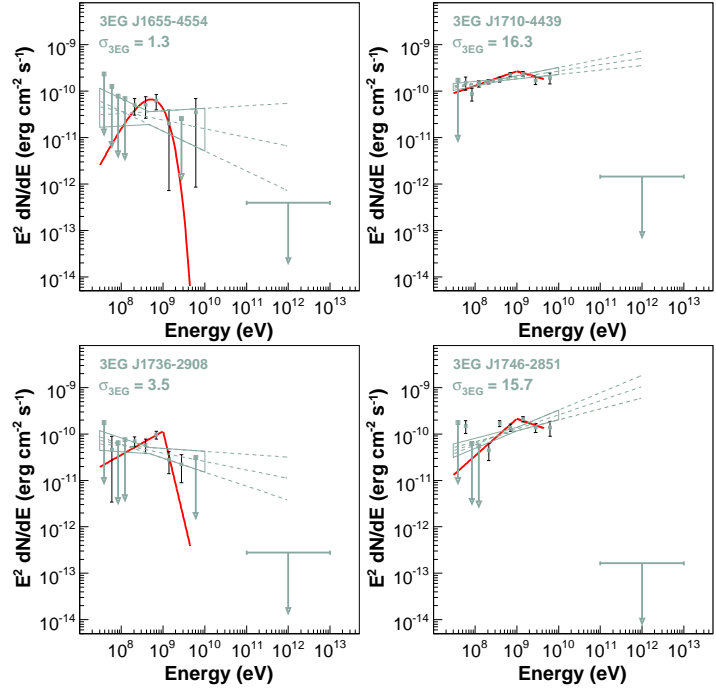} 
  \end{center}
  \caption{SED at E $> 30$ MeV for the non-coincident cases in which
    the EGRET spectrum shows significant deviation from a simple
    power-law form. The previously reported higher order spectral
    representations are shown in red (exponential cutoff for
    3EG\,J1655--4554 and broken power-law for 3EG\,J1710--4439,
    3EG\,J1736--2908 and 3EG\,J1746--2851). }\label{fig::OtherFits}
\end{figure}

The $\chi^2$ of the resulting fits were compared to that for a single
power-law and an F-test employed to test if the more complex form was
justified. For many $\gamma$-ray sources there is insufficient
high-energy data to justify higher order functional fits. However, for
four of the 17 EGRET sources considered in this study the F-test
strongly suggests a different spectral form (with a chance probability
$<0.05$ as discussed in detail in ~\citet{ReimerBertsch01}):
3EG\,J1655--4554 is better fit by a power-law with exponential cutoff,
3EG\,J1710-4439, 3EG\,J1736-2908, and 3EG\,J1746-2851 are best fit
with a broken power-law. All of these sources have no positional
counterpart at TeV energies (and are marked with triangles in
Figure~\ref{fig::EGRETnonconnect}). The different spectral
representations are shown in red in Figure~\ref{fig::OtherFits}. It is
interesting to note, that out of the four sources mentioned above for
which the \HESS\ non-detection strongly suggests a cutoff in the
energy spectrum, the two sources with the largest incompatibility
measure $\sigma_{\mathrm{3EG}}$ are also characterised by a
statistically significant cutoff in the EGRET spectrum. In particular,
the previously mentioned source 3EG\,J1746--2851 (Galactic Centre)
shows strong indications for an energy break below 10~GeV.  The
indicated cutoff in some of the EGRET spectra is entirely consistent
with the constraining VHE limit based on power-law extrapolation. The
prediction that the other two EGRET sources (3EG\,J1809--2328, and
3EG\,J1837--0606) constrained by the \HESS\ upper limits show a cutoff
in the energy range between 10~GeV and 100~GeV is therefore well
justified and will be tested by upcoming GLAST-LAT observations.

\subsection{VHE $\gamma$-ray sources without an EGRET counterpart}
\label{sec::nonconnect::VHE}
In this section the \HESS\ sources without a catalogued EGRET
counterpart are addressed. At all nominal \HESS\ source locations,
flux upper limits have been determined from the EGRET data at energies
above 1~GeV by means of the EGRET likelihood
technique~\citep{Mattox96}. In the determination of the EGRET upper
limit, both the Galactic diffuse emission and point-sources exceeding
a 5~$\sigma$-detection significance threshold were modelled and
subsequently subtracted. The underlying EGRET exposure corresponds to
the first four years of the EGRET mission. As previously discussed,
the sensitivity of EGRET (in terms of energy flux $E^2 dN/dE$) is
considerably worse than the \HESS\ sensitivity so that no EGRET
detection of a \HESS\ source is expected under the assumption of equal
energy flux -- which might obviously not necessarily be fulfilled in
an astrophysical source.

\begin{figure}[ht]
  \begin{center}
    \noindent
    \includegraphics[width=0.85\textwidth]{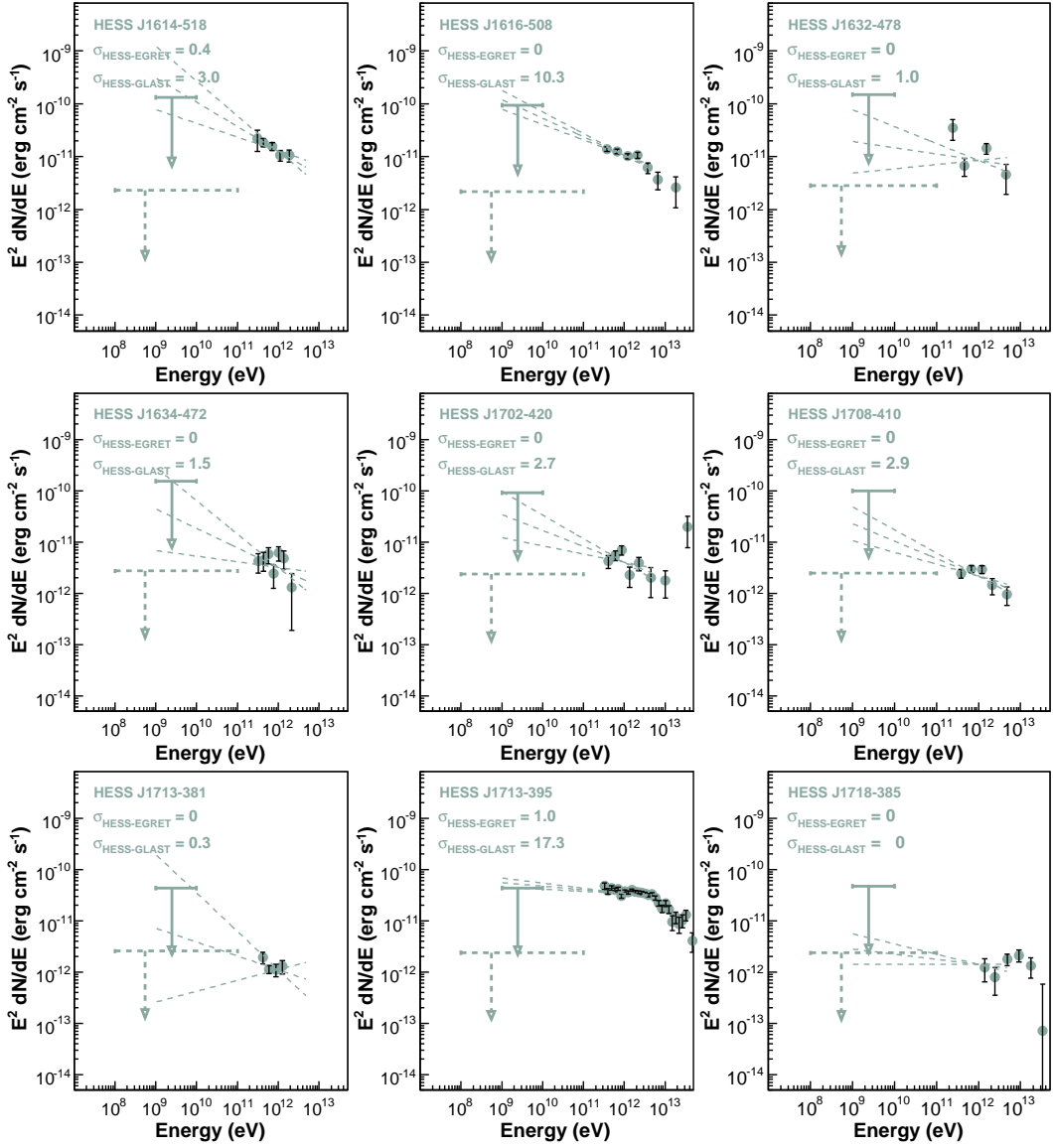} 
  \end{center}
  \caption{(Part 1) SED at E $> 30$ MeV for the cases in which no
    EGRET catalogued counterpart source was found for the \HESS\
    source. The dashed arrow shows the predicted upper limit from a
    one-year GLAST scanning observation, taking into account the
    galactic diffuse background. Derived from this is the spectral
    compatibility parameter $\sigma_{\mathrm{HESS-GLAST}}$ between
    GLAST and H.E.S.S. assuming a non-detection with GLAST to
    illustrate that GLAST will be able to probe the power-law
    extrapolation from VHE $\gamma$-ray energies whereas the existing
    EGRET upper limits are unconstraining in this
    regard.}\label{fig::HESSnonconnectA}
\end{figure}

\begin{figure}[ht]
  \begin{center}
    \noindent
    \includegraphics[width=0.85\textwidth]{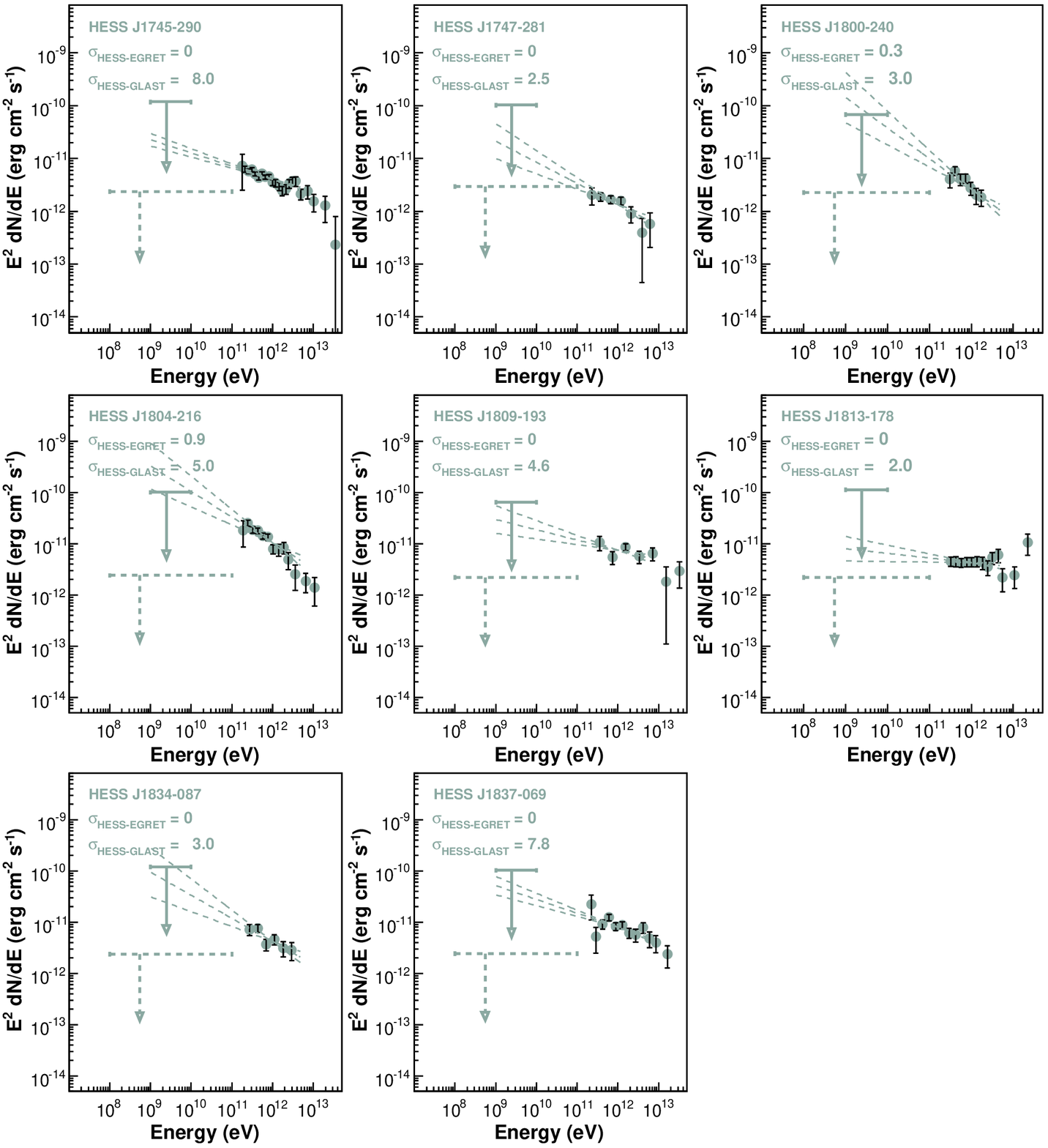} 
  \end{center}
  \caption{(Part 2) SED at E $> 30$ MeV for the cases in which no
    EGRET catalogued counterpart source was found for the \HESS\
    sources.  The dashed arrow shows the predicted upper limit from a
    one-year GLAST scanning observation, taking into account the
    diffuse emission. Derived from this is the spectral mismatch
    between GLAST and H.E.S.S.\ assuming a non-detection with GLAST to
    illustrate the GLAST will be able to probe the power-law
    extrapolation from VHE $\gamma$-ray energies whereas the EGRET
    upper limits are unconstraining in this
    regard.}\label{fig::HESSnonconnectB}
\end{figure}

Methodologically similar to the previous section, the determination of
spectral compatibility was performed by extrapolating \HESS-measured
VHE spectra to 1~GeV and comparing the resulting flux to the EGRET
upper limit at that energy. The spectral compatibility parameter
$\sigma_{\mathrm{H.E.S.S.}}$ is determined in a similar way to
$\sigma_{\mathrm{3EG}}$.  The spectra of \HESS\ sources with
significant curvature were only fitted from the threshold energy at
$\sim 100$~GeV to 1~TeV. As in previous sections,
$\sigma_{\mathrm{HESS-EGRET}}$ describes how well the extrapolated
\HESS\ spectrum can be accommodated by the EGRET upper limit. The
resulting spectral energy distributions (SEDs) of the non-coincident
\HESS\ sources are shown in Figures~\ref{fig::HESSnonconnectA} and
\ref{fig::HESSnonconnectB}.

In all cases, the values of $\sigma_{\mathrm{HESS-EGRET}}$ are less than
or equal to 1, implying that no EGRET upper limit is violated by the
\HESS\ extrapolation to 1~GeV, in stark contrast to the results
discussed in the previous section.  The most interesting case is that
of HESS\,J1713--395 (RX\,J1713.7--3946).  In this case the power-law
extrapolation is at the level of the EGRET upper limit and
$\sigma_{\mathrm{HESS-EGRET}}  = 1$. The unconstraining nature of the
EGRET upper limits is simply a consequence of a lack of instrumental
sensitivity at GeV energies, worsened in regions of pronounced diffuse
$\gamma$-ray emission such as the \HESS\ GPS region. However, this
situation will change significantly in the near future, given the
expected sensitivity of the GLAST-LAT as also shown in
Figures~\ref{fig::HESSnonconnectA} and \ref{fig::HESSnonconnectB} in
which $\sigma_{\mathrm{HESS-GLAST}}$ is calculated for a typical one-year
GLAST sensitivity limit in the Inner Galaxy. These numbers suggest
that the increased sensitivity of the LAT might render common GeV-TeV
studies possible.  While the EGRET upper limits are currently
insensitive to linear extrapolations of the \HESS\ spectra, the
GLAST-LAT will clearly allow for more sensitive studies. It should,
however, be noted, that a linear extrapolation between \HESS\ and
GLAST-LAT energies most probably represents the ``best-case'' for any
such study: physical models typically show spectra that harden towards
GeV energies, unless a different emission component/process takes
over. It remains to be seen if GLAST will detect emission at
comparable energy flux and potentially determine the position of the
peak in the SEDs. As discussed previously, the tremendous advantage of
the GLAST-LAT over any previous mission is the continuous energy
coverage from 30~MeV all the way up into the VHE $\gamma$-ray range at
$\sim 300$ GeV with significantly improved sensitivity and angular
resolution, bridging the current energy gap in which some of the
physically interesting suggested energy cutoffs occur.

\section{Interpretation}
\label{sec::interpretation}

\subsection{Sources detected both at GeV and TeV energies}
\label{subsec::Connection}
As previously stated and shown in Table~\ref{tab::ConnectCases}, only
9 sources exist which can be characterised as coincident Galactic
EGRET and VHE $\gamma$-ray sources at this moment (5 within the inner
Galaxy, 4 outside of the \HESS\ GPS region). Given the large number of
Galactic sources in both GeV and TeV $\gamma$-rays this number is
rather small -- and is indicative of different dominant source classes
in these two energy domains. However, for the few cases where a
positional coincidence may exist some important astrophysical
implications as well as predictions for the upcoming GLAST mission can
be drawn.

Whilst EGRET and in particular GLAST have sufficiently large FoVs to
be able to efficiently observe the whole sky, the limited FoV of
imaging VHE $\gamma$-ray instruments (typically 4$^{\circ}$ diameter)
allow for only limited sky coverage. However, for known GeV sources
high-angular resolution VHE instruments such as MAGIC and \HESS\ with
tremendously higher photon statistics at high energies can help in the
identification and interpretation of the GeV emission. This approach
has been followed by~\citet{Reimer} for the Kookaburra complex. In
this region of TeV and GeV $\gamma$-ray emission, a re-analysis of the
EGRET data taking advantage of the higher spatial resolution images
from \HESS\ observations, demonstrated that the dominant GeV emission
(3EG\,J1420--6038) is positionally coincident with
HESS\,J1420--607~\citep{HESSKooka}. This EGRET source has been flagged
as confused in the 3EG catalogue~\citep{EGRET} and in the re-analysis
3EG\,J1420--6038 was found to be partially overlapping with a less
intense second GeV $\gamma$-ray source. This second GeV source --
detected below the nominal detection threshold for EGRET -- is
apparent in a dedicated analysis at approximately 1/3 of the GeV flux
of the dominant source~\citep{Reimer} and is positionally coincident
with the second VHE $\gamma$-ray source in the Kookaburra region,
HESS\,J1418--609~\citep{HESSKooka} (associated with the ``Rabbit''
PWN). This suggestive morphology match between the GeV data and the
\HESS\ data thus helped in the interpretation and identification of
the confused EGRET sources and made a separation into two individual
sources possible.  Studies such as this one show how confused GeV
emission regions (in particular in the Galactic plane where the
diffuse $\gamma$-ray background is dominant) may be unravelled using
the GeV emission as measured from a large-aperture space-based
$\gamma$-ray instrument together with narrow FoV but superior spatial
resolution observations provided by ground-based atmospheric Cherenkov
telescopes. This approach seems promising for achieving convincing
individual source identifications in the era of the GLAST-LAT.

\begin{figure}[ht]
  \begin{center}
    \noindent
    \includegraphics[width=0.65\textwidth]{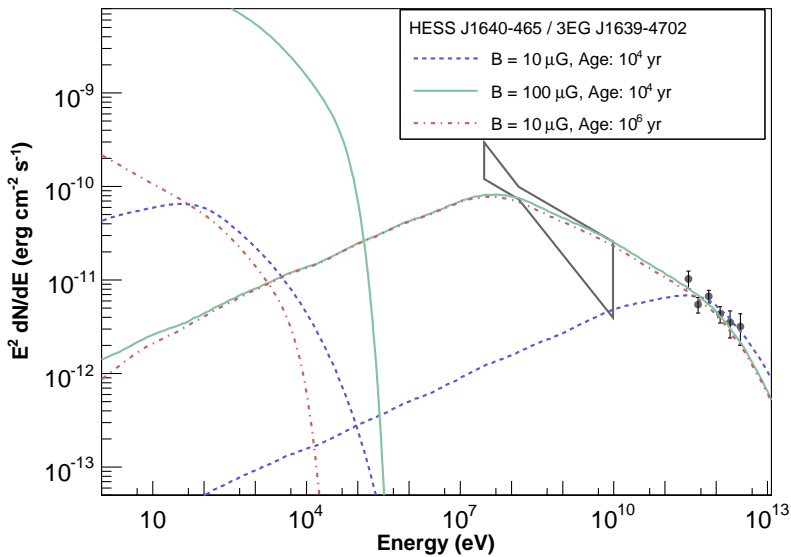} 
  \end{center}
  \caption{SED for the coincident source HESS\,J1640-465 along with
    leptonic IC-models for different magnetic fields and different
    ages of the system. The purpose of this figure is to demonstrate
    that rather extreme values for the magnetic-field ($\sim 100
    \mu$G) or the age of the system ($\sim 10^6$ years) have to be
    invoked to fit such a spectral energy distribution in a leptonic
    model. These models numerically take into account the
    time-evolution of the electron spectrum considering energy losses
    and injection of electrons in time-steps much shorter than the age
    of the system. Synchrotron and IC losses are calculated following
    the formalism in~\citet{Blumenthal}. The injection spectrum for
    the electrons was chosen to have a photon index of 2.5, the
    Inverse Compton scattering was performed on the CMB only.  It
    should be noted, that the X-ray flux between 2 and 10~keV detected
    from this source is at the level of $10^{-13}$ erg cm$^{-2}$
    s$^{-1}$ as determined by~\citet{Funk1640}. This rather low X-ray
    flux renders a connection between the \HESS\ and the EGRET source
    in any leptonic model extremely difficult as demonstrated by this
    figure.}
  \label{fig::HESSConnectModelling}
\end{figure}

On the other hand, the detection of VHE $\gamma$-ray sources with
EGRET (or the GLAST-LAT) may help in the interpretation of the TeV
data and the modelling of the $\gamma$-ray emission mechanism.
Measuring the energy spectrum of a high-energy $\gamma$-ray source
over 5--6 decades in energy should provide rather stringent
constraints on the $\gamma$-ray emission mechanism. The Crab is in
this respect the only good example of a Galactic source for which both
an excellent GeV and TeV coverage exists which in turn helped to
understand the emission mechanism and the magnetic field strength
rather well in comparison to most other $\gamma$-ray sources. With the
advent of the GLAST-LAT many more such sources with a good GeV and TeV
coverage can be expected.

Figure~\ref{fig::HESSConnectModelling} shows the SED for the
positionally coincident sources 3EG\,J1639--4702 and
HESS\,J1640--465~\citep{HESSScanII, Funk1640}. The figure shows a
rather typical $\gamma$-ray SED for a positionally coincident sources
(see Figure~\ref{fig::SpectraConnect}) with a power-law spectrum at
TeV energies with photon index $2.4\pm 0.15$ and a similar power-law
at GeV energies with photon index $2.5 \pm 0.18$, at an energy flux
level an order of magnitude higher than that at 1~TeV. The EGRET
source 3EG\,J1639--4702 is rather close to the detection significance
threshold (with a TS$^{1/2}$-value of 6.4).  Taking this SED as
representative, several scenarios for a common origin of the
$\gamma$-ray emission are considered. For a hadronic model the shape
of this SED can be rather easily fitted, requiring a power-law
distribution of primary hadrons with $dN/dE \propto E^{-\alpha}$, with
$\alpha\approx2.5$ and a maximum particle energy beyond the TeV range.
However, for a simple leptonic model, with the $\gamma$-ray emission
interpreted as inverse-Compton up-scattering of soft photon fields,
matching the shape of the SED requires rather extreme values for the
magnetic field~\citep[given the low level of X-ray synchrotron
emission from this system as reported by][]{Funk1640} or for the age
of the system (given the need to confine the accelerated electrons
within the system).  This is demonstrated in
Figure~\ref{fig::HESSConnectModelling} which shows 3 leptonic model
curves. In the generation of these models, the time-evolution of the
electron spectrum due to energy losses was taken into account. These
energy losses were calculated according to the formalism described
in~\citet{Blumenthal}. For high energy electrons the energy-loss
(cooling) timescale $E/(dE/dt)$ is proportional to $1/E$ for losses
predominantly via synchrotron radiation or IC in the Thomson
regime. In this case, for continuous injection of electrons with a
power law spectrum $dN/dE \propto E^{-\alpha}$, a spectral break to
$E^{-(\alpha+1)}$ will occur. The slope of the IC spectrum (again in
the Thomson regime) is given by $\Gamma=(\alpha+1)/2$. In the
idealised case of the Thomson cross-section and a single (thermal)
target radiation field the break energy is given approximately by:
\begin{equation}
E_{\mathrm{break}} \approx 0.4 (t_{\mathrm{source}}/10^{6}
\mathrm{yr})^{-2}((U_{\rm{rad}}+B^{2}/8\pi)/ 1 \rm{eV\,cm}^{-3})^{-2}
(T/ 2.7\,\rm{K}) \,\rm{GeV} 
\end{equation}
In all cases shown in Figure~\ref{fig::HESSConnectModelling}, the
time-independent injection spectrum of the electrons was fixed with an
index of 2.5 and a cutoff energy at 100~TeV, with the IC scattering on
the cosmic microwave background (CMB) alone.
The first curve (dashed blue) is derived using values rather typically
assumed for TeV sources: a magnetic field strength $10\mu$G and age of
$10^4$ years. This curve provides an adequate description of the
\HESS\ data, but not the EGRET data due to the characteristic turnover
of the $\gamma$-ray spectrum at lower energies. The other two curves
(solid green and dash-dotted red) are shown to illustrate how the SED
could be accommodated in a leptonic model and thus how the peak of the
IC emission can be pushed into the EGRET range. Taking a typical
Galactic radiation field (which might not be realistic as e.g.\ in
binary system with a massive stellar component) either rather high
magnetic fields (green solid) or rather old sources have to be invoked
(dash-dotted red). The high-magnetic field scenario would, however,
lead to the prediction of a very high X-ray flux.  This prediction
contradicts the faint X-ray emission detected from this object (at the
level of $10^{-13}$ erg cm$^{-2}$ s$^{-1}$) as well as in most other
Galactic VHE $\gamma$-ray sources (where the X-ray emission is
typically at the same level or below the VHE $\gamma$-ray energy
flux). To explain the $\gamma$-ray emission of coincident sources
through leptonic IC emission, the sources should thus be rather old to
be able to accumulate enough low energy electrons to explain the high
GeV flux in a typical Galactic radiation field. They should then,
however, either be rather bright X-ray emitters or be very old.

\begin{figure}[ht]
  \begin{center}
    \noindent
    \includegraphics[width=0.65\textwidth]{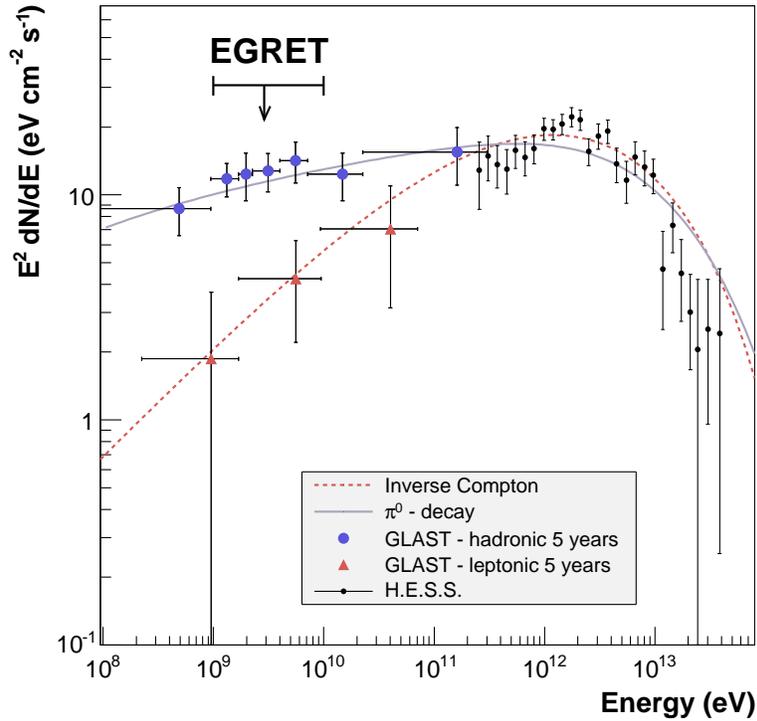} 
  \end{center}
  \caption{High-energy SED for the SNR RX\,J1713.7--3946. The black
  data points show measurements with H.E.S.S., whereas the blue
  circles and red triangles show simulated GLAST data, assuming two
  different models (leptonic and hadronic) for the $\gamma$-ray
  emission (shown as dashed red and solid blue lines). This simulation
  uses the current best estimate of the LAT performance and illustrate
  that in principle the GLAST-LAT should be able to detect this
  prominent shell-type SNR in a 5-years observation or faster,
  depending on the emission mechanism. This figure has been reproduced
  from~\citep{FunkICRCSNR}. }
  \label{fig::SED1713}
\end{figure}

VHE $\gamma$-ray sources may be detectable using GLAST even if the
$\gamma$-ray emission is generated by IC scattering on a typical
Galactic radiation field, as demonstrated for the SNR
RX\,J1713.7--3946 where a GLAST detection should shed light on the
heavily debated origin of the TeV
emission~\citep{FunkICRCSNR}. $\gamma$-rays of leptonic origin
(produced by IC) might be distinguishable from those of hadronic
origin (produced by $\pi^0$-decay) through their characteristic
spectral shape, although recent claims have been made that under
certain conditions the leptonic $\gamma$-ray spectra might resemble
those of pionic decays~\citep{Ellison}. Figure~\ref{fig::SED1713}
shows that the GLAST-LAT will have the sensitivity to measure energy
spectra (in 5 years of scanning observations) for both hadronic and
leptonic emission scenarios, illustrating that the LAT energy range is
particularly well suited to distinguish these models. Measuring the
spectral shape of the $\gamma$-ray emission through deep GeV
observations with the GLAST-LAT will play an important role in
interpreting the currently known TeV $\gamma$-ray sources.

\subsection{The non-connection of GeV and TeV sources}

For sources where no positional coincidence has been found for the GeV
and TeV domains both instrumental and astrophysical explanations can
be invoked.

\subsubsection{Instrumental reasons for non-connection}

The most obvious reason for a non-detection of a TeV source with EGRET
is the sensitivity mismatch. In a typical $\sim 5$ hour observation
\HESS\ has an energy flux sensitivity of about a factor of $\sim
50-80$ lower than that of EGRET for its entire lifetime (above 1~GeV
in the Galactic Plane). Additionally, with decreasing detection
significance an increasing number of EGRET sources are expected to be
artificial due to source confusion in the Galactic plane and in
particular due to uncertainties from the model chosen to describe the
dominant diffuse $\gamma$-ray emission.  The GLAST-LAT will inevitably
shed more light on all persistent EGRET sources, since these will be
rather bright $\gamma$-ray sources for the LAT instrument. However, it
should be noted that the brightest Galactic H.E.S.S.\ sources (such as
RX\,J1713.7--3946) are not going to be very bright GLAST sources as
discussed in the previous section. Certainly, similar to EGRET, the
LAT will (at the lower end of the energy range) suffer from
uncertainties and systematic effects due to intrinsic properties of
the experimental approach and in particular due to the modelling of
the diffuse $\gamma$-ray background, however, at a lower flux level.

Another instrumental effect that could render a correlation between
GeV and TeV sources unlikely, is the insensitivity of imaging VHE
$\gamma$-ray instruments to very extended sources (radius $>
1^{\circ}$) without significant sub-structure. The EGRET data do not
put strong constraints on the source extension of a typical source in
the Galactic plane. Source extensions that can be derived from the
data are on the scale of the EGRET PSF, i.e. degree scales. The
angular resolution (and thus the maximum sensitivity) of VHE
$\gamma$-ray instruments on the other hand is of the order of a few
arc minutes. The upper limits for \HESS\ at the positions of EGRET
sources quoted in this study are derived under the assumption of a
point-like source (with a typical size of the source region of less
than $\sim 0.1^{\circ}$ rms width). The sensitivity and thus the upper
limit scales roughly linearly with the source size~\citep{FunkPhD} and
for source sizes in excess of $\sim 1^{\circ}$, the \HESS\ data become
completely unconstraining due to the fact that the source size becomes
comparable with the size of the FoV and no reliable background
estimation can be performed \citep[see][for a description of the
background estimation techniques used]{BergeBackground}. Large-FoV
instruments (with poorer angular resolution) such as
Milagro~\citep{Milagro}, are better suited to detect sources with
intrinsically large sizes in VHE $\gamma$-rays (with sufficiently high
fluxes). However, due to their modest ($\sim1^{\circ}$) angular
resolution, such instruments suffer from problems of source confusion
similar to those of current GeV measurements. Indeed, several of the
recently reported Milagro source candidates are coincident with EGRET
sources~\citep{MilagroPlane}.  Hypothesising that EGRET sources
exhibit angular sizes larger than $\sim 1^{\circ}$, Milagro-type
instruments might be better suited to detect large scale emission at
VHE $\gamma$-ray energies.  Again, the GLAST-LAT, with its superior
angular resolution to EGRET, will shed more light on the issue of the
intrinsic sizes of GeV sources in the Galactic plane. The constraints
on the power-law extrapolation of EGRET sources by sensitive \HESS\
upper limits as derived in the previous sections are naturally only
valid under the assumption that the VHE counterpart to the EGRET
emission does not exhibit a size much larger than $\sim 1^{\circ}$.

\subsubsection{Astrophysical reasons for non-connection}

The non-detection of most TeV sources in the GeV range by EGRET may be
due simply to a lack of instrumental sensitivity. On the other hand,
the lack of TeV counterparts to most bright GeV sources requires the
presence of steepening (or cut-offs) between 10 and 100 GeV in the
spectra of these sources (see section~\ref{sec::nonconnect} and
Figure~\ref{fig::OtherFits}). Steepening in $\gamma$-ray energy
spectra between 10 and 100~GeV can occur for many reasons, the most
prominent of which are discussed briefly below.

\emph{Acceleration limits}. The maximum energy to which particles are
accelerated in a source may be determined by a balance between the
acceleration and energy loss timescales, or between acceleration and
escape timescales, or simply by the lifetime of the source.  In the
limit of Bohm diffusion, the escape time of accelerated particles from
the source can be written as
\begin{equation}
t_{\mathrm{escape}} \sim (r_{\mathrm{source}}/\mathrm{pc})^{2} D_{0}
(E/\mathrm{TeV})^{-\Delta}
\end{equation}
The associated cut-off in the resulting $\gamma$-ray emission may
occur at much lower energies, as in the case of proton-proton
interactions ~\citep[a factor $\sim20$ as shown in][]{Kappes2007}, or
close to the primary particle energy, as in the case of inverse
Compton scattering in the Klein-Nishina limit~\citep{Blumenthal}.

\emph{Particle transport} may impact on the spectral shape in several
ways. For protons described by a power-law $ J_p(E_p)=K E_p^{-\Gamma}$
the $\gamma$-rays produced in hadronic interactions are expected to
follow a similar power-law spectrum $F_{\gamma}(E_{\gamma})\propto
E_{\gamma}^{-\Gamma}$. Generally, high energy particles escape more
easily leading to a cut-off in the particle and hence $\gamma$-ray
spectrum inside the source. Therefore, due to particle transport, the
spectrum of the protons generating the $\gamma$-rays through hadronic
interactions is not necessarily the same as the one at the
acceleration site. In the case of diffusion the proton spectrum at the
$\gamma$-ray production site can instead be written as
$J_p(E_p,\;r,\;t)=\frac{c}{4\pi} f, $ where $f(E_p,\;r,\;t)$ is the
distribution function of protons at an instant $t$ and distance $r$
from the source. The distribution function satisfies the diffusion
equation~\citep{Ginzburg}.
\begin{equation}\label{difeq}
   \frac{\partial f}{\partial t}=\frac{D(E_p)}{r^2} \frac{\partial}{\partial
   r} r^2 \frac{\partial f}{\partial r} + \frac{\partial}{\partial
   E_p}(Pf)+Q,
\end{equation}
where $P=-dE_p/dt$ is the continuous energy loss rate of the
particles, $Q=Q(E_p,\;r,\;t)$ is the source function, and $D(E_p)$ is
the diffusion coefficient. ~\citet{Atoyan} derived a general solution
for Equation (\ref{difeq}).  Hence, as has been emphasised
by~\citet{FelixSN1006}, the observed $\gamma$-ray flux can have a
significantly different spectrum from that expected from the particle
population at the source. In the (expected) case of energy-dependent
diffusion ($D \propto E^{-\Delta}$, with $\Delta$ typically assumed to
lie in the range $\sim 0.3-1.0$) the $\gamma$-ray spectrum will follow
$F_{\gamma}(E_{\gamma})\propto E_{\gamma}^{-(\Gamma+\Delta)}$. The
exact shape of the spectrum will depend on the age of the accelerator,
duration of injection, the diffusion coefficient, and the location of
the target material.

The influence of convection (lower energy cutoff in primary particle
spectrum) is typically stronger for low energy (GeV) $\gamma$-rays
potentially resulting in a VHE $\gamma$-ray source that has no EGRET
counterpart in cases in which an external accelerator produces primary
hadrons near an active target.\citet{Torres2004} and~\citet{Eva2006}
have recently studied collective wind configurations produced by a
number of massive stars, and obtained densities and expansion
velocities of the stellar wind gas that is the target for hadronic
interactions in several examples, showing that these may be sources
for GLAST and the TeV instruments in non-uniform ways, i.e., with or
without the corresponding counterparts in the other energy band.

\emph{Particle energy losses} away from the acceleration site may also
produce spectral steepening in a very natural way as discussed earlier
(see section~\ref{subsec::Connection}). In the case where particle
injection is effectively finished (i.e. the injection rate is much
lower than in the past), radiative energy losses may produce a rather
sharp cut-off in the $\gamma$-ray spectrum as e.g. shown in
~\citep{Funk1640}. For high energy electrons the energy-loss (cooling)
timescale $E/(dE/dt)$ is proportional to $1/E$ for losses dominantly
via synchrotron radiation or IC in the Thomson regime. In this case,
for continuous injection of electrons with a power law spectrum $dN/dE
\propto E^{-\alpha}$, a spectral break to $E^{-(\alpha+1)}$ will
occur. The slope of the IC spectrum (again in the Thomson regime) is
given by $\Gamma=(\alpha+1)/2$. In the idealised case of the Thomson
cross-section and a single (thermal) target radiation field the break
energy is given approximately by:
\begin{equation}
E_{\mathrm{break}} \approx 0.4 (t_{\mathrm{source}}/10^{6}
\mathrm{yr})^{-2}((U_{\rm{rad}}+B^{2}/8\pi)/ 1 \rm{eV\,cm}^{-3})^{-2} (T/
2.7\,\rm{K}) \,\rm{GeV}
\end{equation}

\emph{$\gamma$-$\gamma$ pair-production} occurs above a threshold
$\epsilon_{\gamma}\epsilon_{\rm{target}}> 2m_{e}^{2}c^{4}$.  For
stellar systems with $\epsilon_{\rm{target}} \sim$ 1 eV, this process
occurs above $\sim$ 500 GeV. Pairs produced in $\gamma$-$\gamma$
interactions may inverse Compton scatter on the same radiation field
-- leading to the development of a
cascade~\citep{Protheroe}. Attenuation on the interstellar IR and CMB
can be neglected below 10 TeV so $\gamma$-$\gamma$ 'cut-offs' are only
expected in compact regions of very high radiation density, for
example within binary stellar systems.  These absorption/cascade
'features' may not represent the end of the $\gamma$-ray spectrum as
emission may recover at energies above the resonance.

\subsection{Prospects for the GLAST-LAT}

As demonstrated by Figures~\ref{fig::HESSnonconnectA}
and~\ref{fig::HESSnonconnectB} the GLAST-LAT should be able to detect
several of the VHE $\gamma$-ray sources in the inner Galaxy, assuming
a simple power-law extrapolation of the spectrum from TeV to GeV
energies. However, this power-law assumption may not be valid in
several cases, as discussed in the following for the known TeV source
classes.

{\emph{Pulsar Wind Nebulae}} are currently the most abundant VHE
$\gamma$-ray sources in the Galactic plane. The most prominent example
is the Crab Nebula~\citep{WhippleCrab, HEGRACrab, MilagroCrab,
HESSCrab, MAGICCrab}. The SED expected of PWNe does not typically
result in significant GeV fluxes: the Crab Nebula, detected throughout
both energy bands, seems to be an exceptional case due to its very
strong magnetic field and relative proximity (2~kpc).  Most VHE
$\gamma$-ray PWNe are expected to be dominated by IC emission for
which the energy flux generally turns down at lower energies. The
position of this inverse Compton peak determines detectability for
both GeV and TeV instruments. Also the size and flux of the source
also obviously affect the detectability with GLAST-LAT.  In general
the higher the energy of the inverse-Compton peak in these sources,
the lower the chance will be to detect them with GLAST.  If a large
fraction of the GeV emission attributed to EGRET Galactic unidentified
sources is related to pulsed magnetospheric emission from pulsars as
opposed to emission from the extended wind nebula then a correlation
between the H.E.S.S.\ and EGRET sources in the Inner Galaxy could be
expected, given that the majority of the H.E.S.S.\ sources in this
region seem to be PWNe associated with energetic
pulsars~\citep{SvenjaPulsars}. However, this expectation may not hold
in general due to diversity of parameters like the beaming geometry or
different conversion efficiency of the pulsar's spin-down power into
the Nebula and into $\gamma$-rays.

{\emph{Shell-type Supernova remnants}} The two prominent and bright
VHE $\gamma$-ray SNRs (RX\,J1713.7--3946 and RX\,J0852.0--4622) are
not expected to be very bright GLAST-LAT sources. Nevertheless, they
are probably amongst the more easily detectable TeV sources in the
GLAST-LAT band.  A detailed simulation of the expected signal from
RX\,J1713.7--3946 shows that it might be detectable in one year of
GLAST-LAT observations depending on the assumed TeV $\gamma$-ray
emission mechanism as shown in the previous section. Morphological
studies in GeV $\gamma$-rays will either have to struggle with
moderate angular resolution at low energies or with low photon
statistic at high energies. However, spectral studies will be
immediately possible following a potential detection as shown in
Figure~\ref{fig::SED1713} for RX\,J1713.7--3946. For RX\,J0852.0--4622
(Vela Junior) the situation is even further complicated by the
close-by bright Vela Pulsar. While both of these prominent
TeV-emitting objects are rather young ($\sim 2000$ years), there is
the potential of older SNRs acting as stronger GeV emitters (but
rather faint TeV sources). In this case the GLAST-LAT might see a
different population of shell-type SNRs than VHE $\gamma$-ray
instruments, namely older SNRs which have accumulated a large number
of lower energy CRs, but for which the higher energy CRs (those that
may give rise to the TeV emission) have already left the acceleration
site. A common detection both with GLAST and VHE $\gamma$-ray
instruments might require a hadronic origin of the $\gamma$-ray
emission rather than an Inverse compton (IC) origin due to the
characteristic turn-over of the IC spectrum at lower energies.

{\emph{Gamma-ray Binary systems}} host a variety of non-thermal
phenomena.  The TeV detected binaries: LS\,5039~\citep{HESSLS5039II},
PSR\,B1259-63~\citep{HESS1259}, LSI\,+61\,303~\citep{MAGICLSI} and
Cyg\,X--1~\citep{MAGICCygX} are currently seen as candidates for
detection at GeV energies. $\gamma$-$\gamma$ absorption in binary
systems may producing anti-correlation of the TeV to GeV radiation
during the orbit of these systems. These orbital modulations are
predicted in basically all models for these systems, irrespective of
the assumptions of a pulsar or a black hole compact object or the
process by which high-energy radiation is emitted~\citep[see
e.g.][]{Dermer2007, Dubus2006, Paredes2006}. Details in predicted
light-curves and spectral evolution in time are however rather
distinctive~\citep{Felix5039, Diego5039}.

\section{Summary}

The main results of the study of the relationship between GeV and TeV
sources are:

\begin{enumerate}
\item There are rather few spatially coincident GeV-TeV sources for
  the considered Galactic region.
\item Those few positional coincident GeV-TeV sources could occur by
  chance, the chance probability of detecting two coincident sources
  within the \HESS\ GPS region is $\sim 40$\%, thus no strong hint for
  a common GeV/TeV source population is detected.
\item Spectral compatibility (based on a power-law extrapolation) seems
  present for most of the positionally coincident sources, but again,
  this is expected to occur by chance (as described in the text) given
  the sensitivity mismatch and the different energy bands.
\item Dedicated \HESS\ limits at the position of the EGRET sources are
  constraining for a power-law extrapolation from the GeV to the TeV
  range for several of the EGRET sources, strongly suggesting cutoffs in
  the energy spectra of these EGRET sources in the unexplored region
  below 100~GeV. Power-law extrapolation of EGRET spectra seem to be
  ruled out for most of the EGRET sources investigated in this study.
\item Dedicated EGRET limits at the position of the \HESS\ sources are
  not constraining for a power-law extrapolation from the TeV to the
  GeV range. This picture will dramatically change once the GLAST-LAT
  with its improved sensitivity over EGRET is in orbit.
\item Several important mechanisms for cutoffs in the energy spectra
  of GeV sources have been discussed. There are well motivated
  physical reasons why the population of GeV and of TeV sources might
  be distinct.
\item If a source can be detected with both GeV and TeV instruments,
  the huge energy ``lever arm'' over 5-6 decades in energy will
  undoubtedly provide stringent constraints on the $\gamma$-ray
  emission mechanism in these Galactic particle accelerators.
\end{enumerate}

Summarising, the study presented here shows that the GLAST-LAT will
tremendously advance the study of the relationship between GeV and TeV
sources by improving the sensitivity over EGRET by an order of
magnitude and in particular by bridging the currently uncovered energy
range between 10~GeV and 100~GeV.

\acknowledgments The authors would like to acknowledge the support of
their host institutions. In particular, S.F.\ acknowledges support of
the Department of Energy (DOE) and Stanford University and would like
to thank the whole \HESS\ and the GLAST-LAT collaborations for their
support and helpful discussions on the topic, in particular Werner
Hofmann, Felix Aharonian, Benoit Lott, and Seth Digel. DFT is
supported by the Spanish MEC grant AYA 2006-00530, he acknowledges
Juan Cortina and other members of the MAGIC collaboration for advice
and encouragement. OR is supported by the National Aeronautics and
Space Administration under contract NAS5-00147 with Stanford
University. JAH is supported by an STFC Advanced Fellowship.

\end{document}